\documentclass{aa}
\usepackage{graphicx}
\usepackage{txfonts}
\defcitealias{2023A&A...675A.130M}{Paper I}

\usepackage{natbib,twoopt}
\usepackage[breaklinks=true, colorlinks=true,urlcolor=blue,citecolor=blue,linkcolor=blue]{hyperref} 
\bibpunct{(}{)}{;}{a}{}{,}             
\makeatletter
  \newcommandtwoopt{\citeads}[3][][]{\href{http://adsabs.harvard.edu/abs/#3}%
    {\def\hyper@linkstart##1##2{}%
     \let\hyper@linkend\@empty\citealp[#1][#2]{#3}}}
  \newcommandtwoopt{\citepads}[3][][]{\href{http://adsabs.harvard.edu/abs/#3}%
    {\def\hyper@linkstart##1##2{}%
     \let\hyper@linkend\@empty\citep[#1][#2]{#3}}}
  \newcommandtwoopt{\citetads}[3][][]{\href{http://adsabs.harvard.edu/abs/#3}%
    {\def\hyper@linkstart##1##2{}%
     \let\hyper@linkend\@empty\citet[#1][#2]{#3}}}
  \newcommandtwoopt{\citeyearads}[3][][]%
    {\href{http://adsabs.harvard.edu/abs/#3}
    {\def\hyper@linkstart##1##2{}%
     \let\hyper@linkend\@empty\citeyear[#1][#2]{#3}}}
\makeatother

\begin{document}

\title{Comparative clustering analysis of \ion{Ca}{ii} 854.2 nm spectral profiles from simulations and observations}

\author{Thore E. Moe
	\inst{1,2}
	\and Tiago M.D. Pereira
	\inst{1,2}
	\and Luc Rouppe van der Voort
	\inst{1,2}
	\and Mats Carlsson
	\inst{1,2}
	\and Viggo Hansteen
	\inst{1,2,3,4}
	\and Flavio Calvo
	\inst{5}
	\and Jorrit Leenaarts
	\inst{5}
	}
\institute{Rosseland Centre for Solar Physics, University of Oslo, P.O. Box 1029 Blindern, NO--0315 Oslo, Norway
\and
Institute of Theoretical Astrophysics, University of Oslo, P.O. Box 1029 Blindern, NO--0315 Oslo, Norway
\and
Lockheed Martin Solar \& Astrophysics Laboratory, 3251 Hanover St., Palo Alto, CA 94304, USA
\and
Bay Area Environmental Research Institute, NASA Research Park, Moffett Field, CA 94035, USA
\and
Institute for Solar Physics, Dept. of Astronomy, Stockholm University, AlbaNova University Centre, 10691 Stockholm, Sweden} 
\date{}

\abstract 
{Synthetic spectra from 3D models of the solar atmosphere have become increasingly successful in reproducing observations, but there are still some outstanding discrepancies for chromospheric spectral lines, such as \ion{Ca}{ii} and \ion{Mg}{ii}, particularly regarding the width of the line cores. It has been demonstrated that using sufficiently high spatial resolution in the simulations significantly diminishes the differences in width between the mean spectra in observations and simulations, but a detailed investigation into how this impacts subgroups of individual profiles is currently lacking.} 
{We aim to compare and contrast the typical shapes of synthetic \ion{Ca}{ii} 854.2 nm spectra found in Bifrost simulations having different magnetic activity  with the spectral shapes found in a quiet Sun observation from the Swedish 1-m Solar Telescope (SST).} 
{We use clustering techniques to extract the typical \ion{Ca}{ii} 854.2 nm profile shapes synthesized from Bifrost  simulations with varying amounts of magnetic activity. We degrade the synthetic profiles to observational conditions and repeat the clustering, and we compare our synthetic results with actual observations. Subsequently we examine the atmospheric structures in our models for some select sets of clusters, with the intention of uncovering why they do or do not resemble actual observations.} 
{While the mean spectra for our high resolution simulations compare reasonably well with the observations, we find that there are considerable differences between the clusters of observed and synthetic intensity profiles, even after the synthetic profiles have been degraded to match observational conditions. The typical absorption profiles from the simulations are both narrower and display a steeper transition from the inner wings to the line core. Furthermore, even in our most quiescent simulation we find a far larger fraction of profiles with local emission around the core, or other exotic profile shapes, than in the quiet Sun observations. Looking into the atmospheric structure for a selected set of synthetic clusters, we find distinct differences in the temperature stratification for the clusters most and least similar to the observations. The narrow and steep profiles are associated with  either weak gradients in temperature, or temperatures rising to a local maximum in the line wing forming region before sinking to a minimum in the line core forming region. The profiles that display less steep transitions show extended temperature gradients that are steeper in the range $-3 \lesssim \log\tau_{5000} \lesssim -1$.} 
{} 


\maketitle

\section{Introduction}
\label{sec:intro}

Thanks to advances in instrumentation, it is now possible to routinely obtain spatially-resolved spectra from the solar surface at spatial resolutions in the tenths of arcseconds or even finer. These detailed spectra contain a wealth of information. For lines formed in the dynamic solar chromosphere, spectral shapes become increasingly complex, especially as the spatial resolution increases. To extract the maximum information from observations, it is crucial to understand how different spectral lines are formed in a dynamic atmosphere. Three-dimensional, radiative magnetohydrodynamic (3D rMHD) simulations of the solar atmosphere \citep[e.g.][]{1998ApJ...499..914S, 2004A&A...421..741V, 2011A&A...531A.154G, 2015ApJ...812L..30I, 2018A&A...618A..87K, 2022A&A...664A..91P,2023ApJ...944..131H} have become a powerful tool to help interpret spectral observations and learn how spectral lines form.

For the solar photosphere, the self-consistent treatment of convection in 3D simulations can reproduce the (mean) shapes of spectral lines in great detail \citep[e.g.][]{2000A&A...359..729A}, and also leads to a mean temperature stratification that agrees very well with a wealth of observational diagnostics \citep{2013A&A...554A.118P}. However, the solar chromosphere is a much more demanding problem, and current simulations do not yet reproduce the variations in chromospheric lines as well as they do for photospheric lines. For example, synthetic profiles of chromospheric lines tend to be narrower than observed \citep{2009ApJ...694L.128L} and can show weaker emission \citep{2013ApJ...772...90L, 2015ApJ...811...81R}. Despite not yet reproducing all chromospheric line shapes, 3D rMHD simulations have been instrumental in forward modeling studies that shape our understanding of most spatially-resolved line formation, e.g. from the formation of the H$\alpha$ line \citep{2012ApJ...749..136L}, and the \ion{Ca}{ii} H\&K lines \citep{2018A&A...611A..62B}. Such studies are also very important in the development of, and interpretation of data from,new observatories, as shown by the MUSE mission \citep{2022ApJ...926...52D, 2022ApJ...926...53C}, and by the IRIS mission \citep{2014SoPh..289.2733D}, for which a series of papers \citep{2013ApJ...772...89L, 2013ApJ...772...90L, 2013ApJ...778..143P, 2015ApJ...806...14P, 2015ApJ...811...80R, 2015ApJ...811...81R,2015ApJ...813...34L,2015ApJ...814...70R,2017ApJ...846...40L} provides unique insight into the formation of UV lines. 

Most forward modeling studies follow a well-tested pattern of synthesizing spectra from 3D rMHD simulations, using either fully 3D radiative transfer or the 1.5D approximation where each simulation column is treated as an independent plane parallel atmosphere, and then comparing spectral signatures with the thermodynamical conditions of the underlying atmosphere. Given the sheer amount of individual spectral (typically on the order of millions for one simulation), it is not possible to study each spectrum in detail. Most studies so far have focused on either the properties of spatially-averaged spectra, or on the distributions of simple spectral properties (e.g. line shifts, line widths, position and amplitude of emission peaks, etc.). The main goal of this work is to extend previous approaches and use more information from the line profiles by the means of clustering techniques.

In a previous paper \citep[][hereafter \citetalias{2023A&A...675A.130M}]{2023A&A...675A.130M}, we discuss and demonstrate the use of clustering techniques such as $k$-means \citep{1956_Steinhaus,1967_Macqueen} and $k$-Shape \citep{10.1145/2723372.2737793} in a forward modeling context. Through the use of $k$-means and $k$-Shape clustering, we investigated the variety of \ion{Ca}{ii} \mbox{854.2 nm} spectral shapes present in a 3D rMHD simulation, as well as how those shapes correlated with the structure of the atmospheric columns they arose from. Here, we want to extend this approach to different types of atmospheres and observations. Again, we will restrict the analysis to the \ion{Ca}{ii} \mbox{854.2 nm} line, which is a widely observed diagnostic of the chromosphere \citep[e.g.][]{2008A&A...480..515C,2013SoPh..288...89C,2015ApJ...810..145D,2016MNRAS.459.3363Q,2017ApJ...846....9K,2021ApJ...920..125M}. 

In this work, we investigate what are the typical shapes of \ion{Ca}{ii} \mbox{854.2 nm} line profiles, and what they tell us about the solar atmosphere. We study how profile shapes vary across simulations with different amounts of magnetic field. In addition, we make a critical comparison between the synthetic and observed clusters of line profiles. With access to the full thermodynamical state of the underlying simulated atmospheres, we investigate how different clusters of atmospheres are structured, and how different quantities influence the formation of the \ion{Ca}{ii} \mbox{854.2 nm} line.

This paper is organised as follows. In Sect.~\ref{sec:methods} we describe the simulations, spectral synthesis, observations, and clustering methods used. In Sect.~\ref{sec:results} we describe the results from the spectral clustering, look in detail at typical families of spectra, and compare simulations with observations. We discuss our results in Sect.~\ref{sec:discussion} and finish with our conclusions in Sect.~\ref{sec:conclusion}.

\section{Methods}
\label{sec:methods}

\subsection{Simulations}
\label{sec:sims}

We make use of three distinct 3D rMHD simulations run with the Bifrost \citep{2011A&A...531A.154G} code. The goal was not to reproduce exactly the observed region, but to experiment with different amounts of magnetic activity. It should be noted that none of these simulations account for non-equilibrium ionization of Hydrogen.

The first simulation, hereafter \textit{ch012023}, is magnetically quiet and has field configuration resembling a coronal hole. It is the same simulation used by \citetalias{2023A&A...675A.130M}, and is described in more detail by \citet{2022A&A...662A..80M}. Its box size is $12\times12$~Mm$^2$ horizontally (with 23~km horizontal grid size) and 12.5~Mm vertically, and its mean unsigned magnetic field at $z=0$ is 3.7~mT (37~G). The vertical grid is non-uniform, and spread over 512 points.

The second simulation, hereafter \textit{nw012023}, has the same physical extent and horizontal spatial resolution of \textit{ch012023}, but a different magnetic field configuration. Here, stronger magnetic field has been injected into the middle of the box, separating regions of opposite magnetic polarity. Its mean unsigned magnetic field at $z=0$ is 8.6~mT (86~G). The vertical grid spans 16.8 Mm with 824 non-uniformly distributed grid points.

The third simulation, hereafter \textit{nw072100}, is very different from the other two. It has a much larger spatial extent, $72\times72$~Mm$^2$ horizontally, and nearly 60~Mm in the vertical direction. The vertical grid spans 1116 non-uniformly grid points. \citet{2023ApJ...944..131H} describe this simulation in detail. It has regions with much stronger magnetic field, and is included in this study as a more extreme case. It is not meant to reproduce the quiet observations we describe below, but instead as a case study for line profiles in a more active atmosphere. Its spatial resolution, with a horizontal grid size of 100~km, is also coarser than the other two models. As \citet{2023ApJ...944..131H} note, the numerical resolution can also affect the mean spectral properties such as the width, so one should keep that in mind when comparing \textit{nw072100} with the other simulations.

We note that throughout this paper, we define the positive vertical axis to be pointing outwards, i.e. positive vertical velocities correspond to upflows.

\subsection{Synthesizing profiles}
\label{sec:synthesis}

As in \citetalias{2023A&A...675A.130M}, we use the fully 3D radiative transfer code Multi3D \citep{2009ASPC..415...87L}, with the polarization-capable extension (Calvo \& Leenaarts, in prep), to generate synthetic spectra of the \ion{Ca}{ii} 854.2 nm line. Although we are focusing our analysis on the shapes of the intensity profiles, Stokes I, we have computed full Stokes profiles accounting for the Zeeman effect under the field-free approximation (i.e. polarization is accounted for in the final formal solution, using atomic populations that have been iterated to convergence considering only the intensity). Multi3D solves the non local thermodynamical equilibrium (NLTE) radiative transfer problem considering one atomic species at a time, i.e. it does not simultaneously solve for multiple species. Here, we use a model Ca atom which consists of six levels (five bound levels and one continuum level).As 3D radiative transfer is expensive, we have trimmed away the deeper and higher parts of the snapshots, which should have negligible influence on the emergent spectra, in order to speed up computations. We cut the top, a few grid points above the horizontal plane where all simulation columns had exceeded 50 kK, and we cut the bottom, below the horizontal plane where the granulation pattern no longer was discernible in maps of the temperature. This cutting reduces \textit{ch012023} to 410 vertical grid points between 8.0 Mm and $-0.42$ Mm, \textit{nw012023} reduces to 552 vertical grid points between 6.8 Mm and $-1.0$ Mm, \textit{nw072100} to 720 vertical grid points between 29 Mm and $-1.3$ Mm. All spectra have been computed with the assumption of complete redistribution (CRD), which is a reasonable choice for this line \citep{1989A&A...213..360U, 2018A&A...611A..62B}, and we have not accounted for isotopic splitting, which does have some influence on the line shapes \citep{2014ApJ...784L..17L}. Furthermore, statistical equilibrium (SE) was assumed, as non-equilibrium ionization of Ca is unimportant for the formation of the 854.2 nm line \citep{2011A&A...528A...1W}.

Our analysis in this paper is focused on the spectral range $\lambda_0 \pm 0.1 ~\mathrm{nm}$, where $\lambda_0$ is the central wavelength of the \ion{Ca}{ii} 854.2 nm line. This range encompasses the chromospheric line core, as well as parts of the photospheric wings. In terms of formation heights, we find, for all our simulations, that the line core reaches unity optical depth at around $\log\tau_{5000} \approx -5.3$ on average, where $\tau_{5000}$ is the optical depth for light at 500 nm (5000 Å), while the farthest parts of the wings in this spectral range reach unity optical depth at $\log\tau_{5000} \approx -1.2$ on average. The formation height initially increases slowly from the far wings towards the line core, until the transition point from wing to core is reached (at about $\log\tau_{5000} \approx -2)$; from there the formation height rapidly increases towards the maximum at the line core. This is in reasonable agreement with the study by \citet{2016MNRAS.459.3363Q}, which looked at the response functions for the \ion{Ca}{ii} 854.2 nm line in the semi-empirical FALC atmosphere \citep{1993ApJ...406..319F}. 

\subsection{Observations}
\label{sec:obs}

\begin{figure}
        \centering 
                \includegraphics{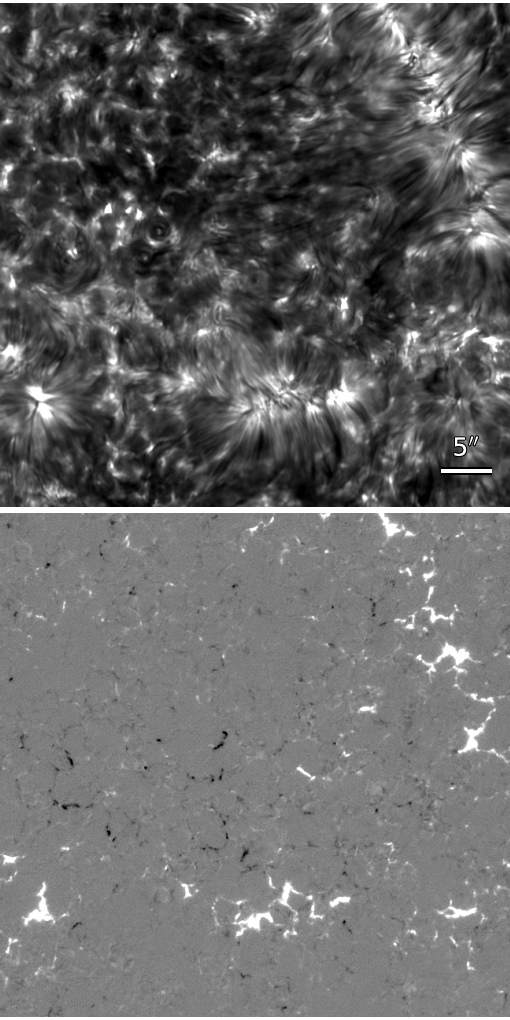}
                        \caption{Overview of the observed region. The images show the field of view used for the spectral clustering, at 2023-06-24T09:15. \emph{Top:} \ion{Ca}{ii} 854.2~nm line core intensity. \emph{Bottom:} \ion{Fe}{i} 630.2~nm line wing Stokes V, a proxy for magnetic field.}
                        \label{fig:obs_line_core}
\end{figure}

The observations were acquired with the CRISP instrument
\citep{2008ApJ...689L..69S} at the Swedish 1-m Solar Telescope \citep[SST,][]{2003SPIE.4853..341S}. CRISP is a Fabry-P{\'e}rot tunable filtergraph that is capable of fast wavelength switching and imaging at high spatial resolution. We observed an area near the edge of an equatorial coronal hole close to disk center at $(x,y) \approx (-119\arcsec,-106\arcsec)$
on 24 June 2014. The heliocentric viewing angle was $\mu \approx 0.99$. CRISP was running a program observing the H$\alpha$ and \ion{Ca}{ii}~854.2~nm lines in spectral imaging mode plus single wavelength \ion{Fe}{i}~630.2~nm spectropolarimetry to produce magnetograms based on \mbox{Stokes V} of the line wing. Here we concentrate on the \ion{Ca}{ii}~854.2~nm data which consist of spectral line scans at 25 wavelength positions between $\pm$0.12~nm with 0.01~nm steps. The full time series started at 08:27:14~UT, has a duration of 01:15:37 and a temporal cadence of 11.5~s (the time it takes to sample the same wavelength again). The data were processed using the CRISPRED data reduction pipeline \citep{2015A&A...573A..40D} which includes multi-object multi-frame blind deconvolution \citep[MOMFBD,][]{2005SoPh..228..191V} image restoration. The seeing conditions were excellent and with the aid of the adaptive optics system and MOMFBD image restoration, the spatial resolution was close to the diffraction limit of the telescope for a large fraction of the time series ($\lambda/D = 0\farcs18$ at the wavelength of \ion{Ca}{ii}~854.2~nm for the $D=0.97$~m clear aperture of the SST). For most of our analysis we use a single time step with particularly good seeing conditions, when the Fried's parameter $r_0$ for the ground-layer seeing was measured to be above 40~cm. The field of view was cropped to about $50\arcsec \times 50\arcsec$ and the plate scale is 0\farcs057~pixel$^{-1}$. In Fig. \ref{fig:obs_line_core} we show an overview of the field of view used for the spectral clustering, including the 854.2~nm line core intensity and a magnetogram from the Stokes V of the \ion{Fe}{i} 630.2~nm line.

\subsection{Degrading the synthetic profiles}
\label{sec:degrading}

In order to fairly compare the simulations to the observations, we need to degrade them spectrally and spatially, as well as resample them. This is done in a four-step process. First the synthetic spectra are convolved with a Gaussian in the spectral domain, using the 10.5~pm full-width-half-maximum (FWHM) spectral instrumental profile of CRISP. Secondly the spectra are downsampled in the spectral domain to match the 21 wavelength points of the narrowband filter, ranging from $\lambda_0 \pm 0.1 ~\mathrm{nm}$, where $\lambda_0$ is the central wavelength of the \ion{Ca}{ii} 854.2 nm line. Thirdly, the spectra are convolved in the spatial domain with a 2D Gaussian with a $0\farcs18$ FWHM to match the telescope's resolution. Finally, the synthetic spectra are interpolated and resampled to match the 0\farcs057~pixel$^{-1}$ plate scale. We note that the synthetic profiles are computed for a disk-center viewing angle, i.e. for $\mu = 1$, and we do not project them to the $\mu \approx 0.99$ viewing angle of the observations because the difference in viewing angle is so minor.

An additional difference between the synthetic and observed spectra is that the synthetic spectra are an instant snapshot of the atmosphere at a given time, while CRISP observations have a given exposure time and scan time (not all wavelengths are observed at the same time), during which time the atmosphere can change. As \citet{2023A&A...669A..78S} show, this should be accounted for in order to do the most accurate comparisons between synthetic and real observables. We did not perform an accurate time-averaged comparison because of several factors: first, the simulation snapshots are typically not saved in such high cadence; second, 3D NLTE radiative transfer is computationally very expensive; and lastly, because the time to acquire each of our CRISP scans is already short (less than 10~s), we do not expect the observed profiles to differ significantly from an `instant' snapshot.

In most of our analysis we use the original, non-degraded synthetic profiles. This is especially relevant when comparing synthetic profiles and atmospheric quantities, since the simulations are not degraded. However, the degraded profiles are an important check both when comparing directly with observations, and also to make sure that the overall range of synthetic spectral clusters is not significantly changed by the observational conditions.

\subsection{Clustering methods}
\label{sec:clustering}

\begin{figure}
        \centering
                \includegraphics[width=0.49\textwidth]{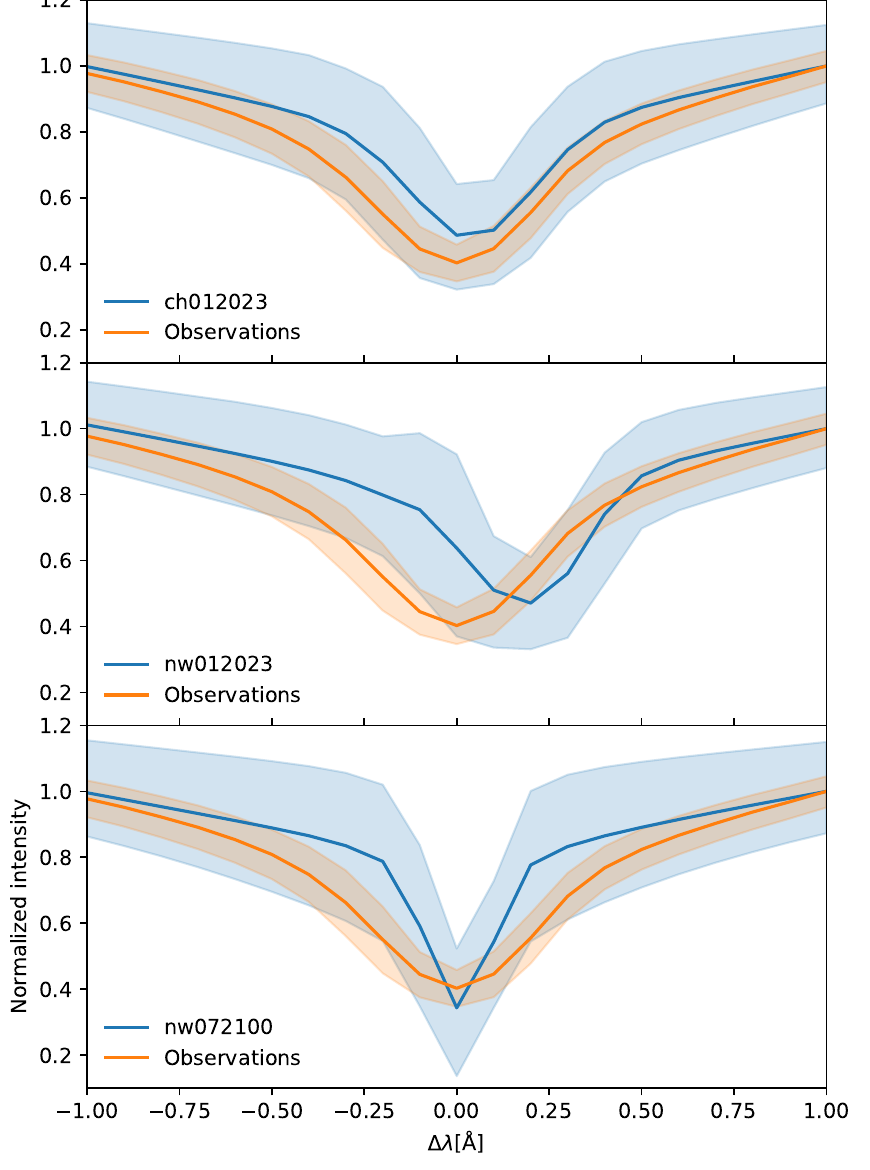}

                        \caption{Mean spectra with 1-$\sigma$ variations for observations and three simulations. The shaded bands show, for each wavelength, the 1-$\sigma$ range of departures from the mean. All spectra were normalized to the local continuum at $\lambda_0+0.1$~nm. The synthetic spectra were degraded to match the conditions of the observations.}
                        \label{fig:mean_spectra}
\end{figure}

\begin{figure*}
        \centering \hspace*{-5mm}
                \includegraphics[width=19.5cm]{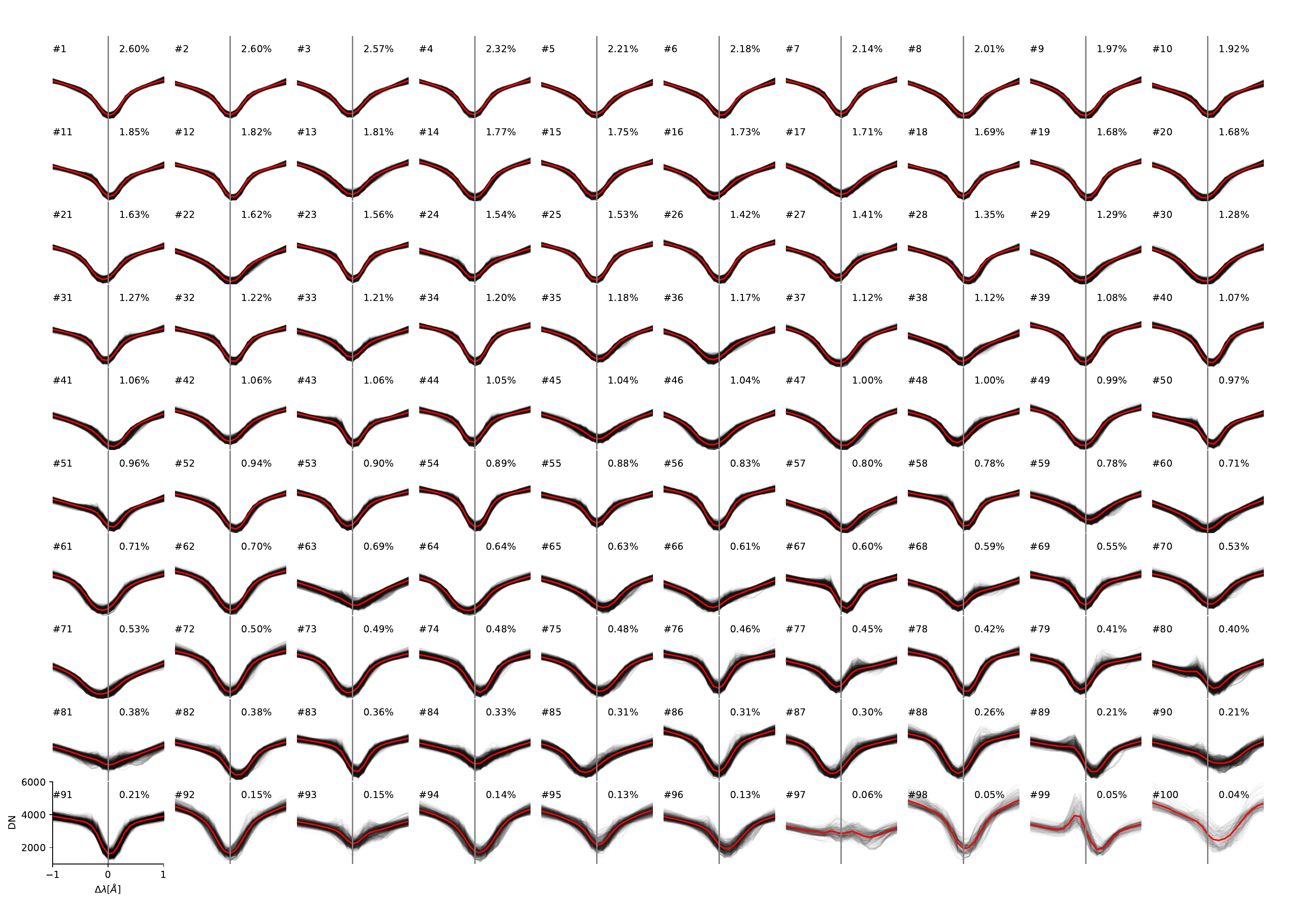}

                        \caption{$k$-means clusters of \ion{Ca}{II} 854.2~nm observed spectra. Using 100 clusters, sorted from most to least frequent. The red line denotes the average of all line profiles belonging to each cluster (show as thin black lines). The fraction of all profiles belonging to each cluster is indicated as a percentage next to the cluster number. The grey line indicates the position of $\lambda_0$, the rest wavelength.}
                        \label{fig:obs250_unscaled_clusters_km}
\end{figure*}

\begin{figure*}
        \centering \hspace*{-5mm}
                \includegraphics[width=19.5cm]{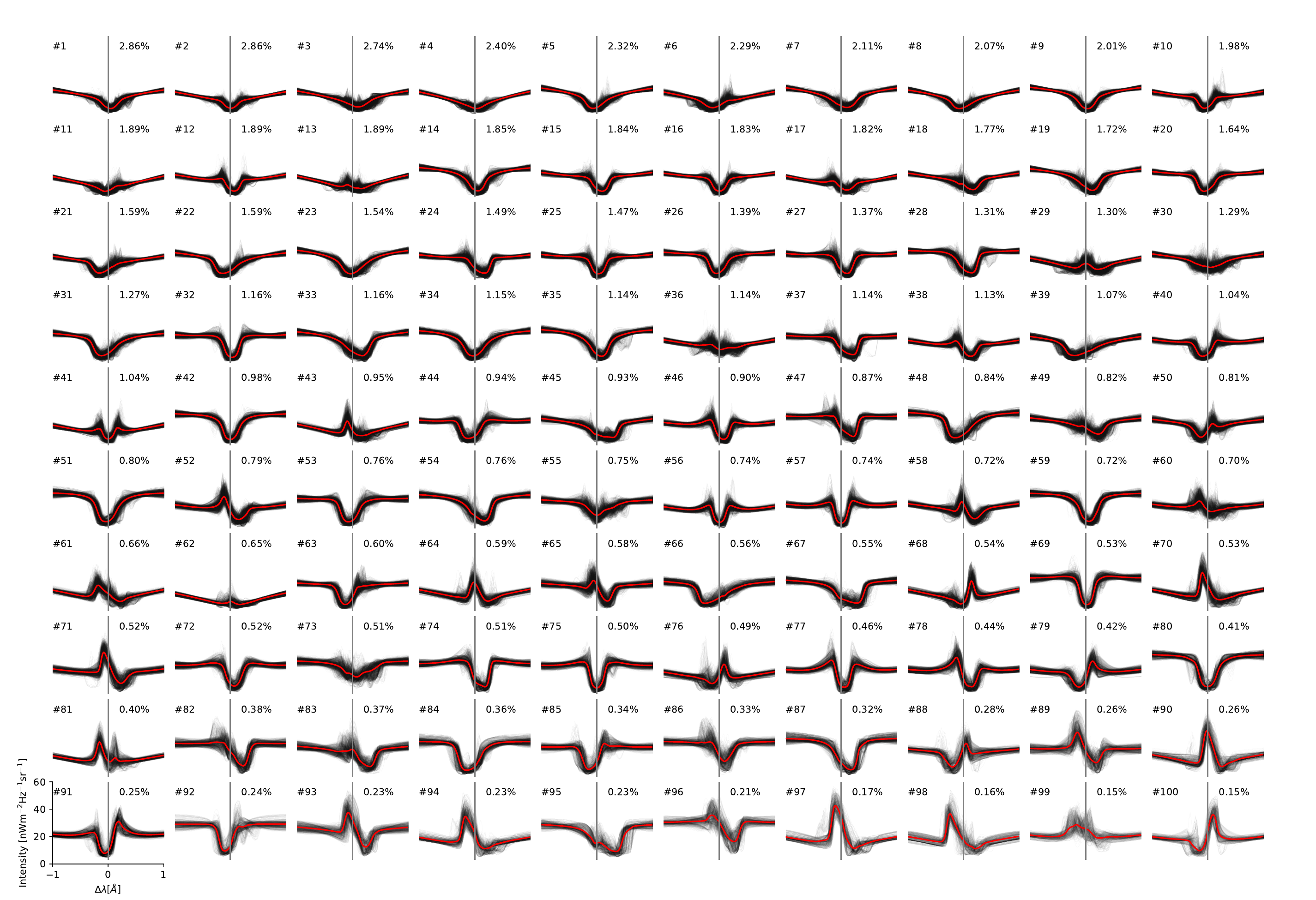}

                        \caption{$k$-means clusters of \ion{Ca}{II} 854.2~nm synthetic spectra from the \textit{ch012023} simulation. The legend is the same as for Fig.~\ref{fig:obs250_unscaled_clusters_km}.}
                        \label{fig:ch012_unscaled_undeg_clusters_km}
\end{figure*}

\begin{figure*}
        \centering \hspace*{-5mm}
                \includegraphics[width=19.5cm]{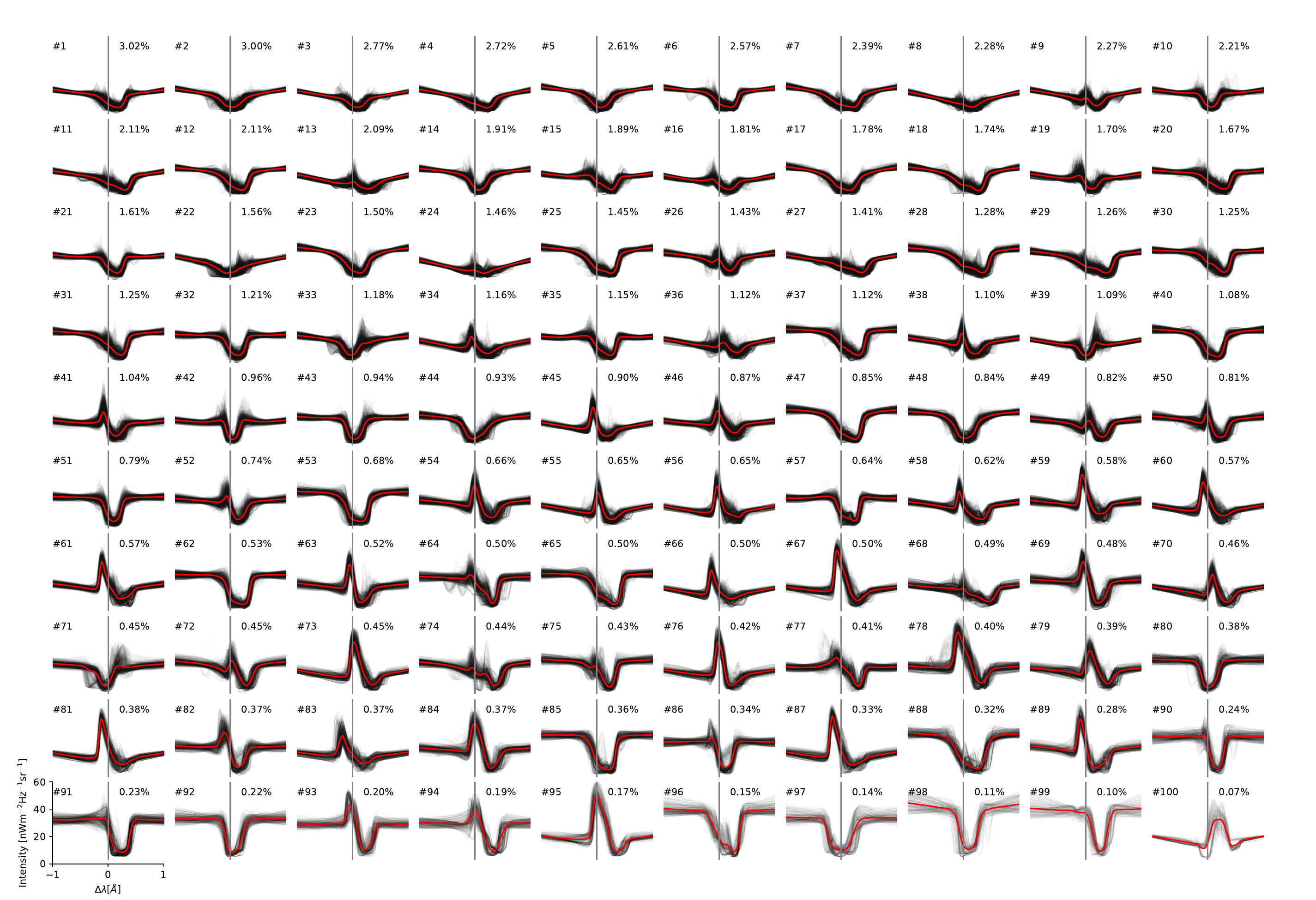}

                        \caption{$k$-means clusters of \ion{Ca}{II} 854.2~nm synthetic spectra from the \textit{nw012023} simulation. The legend is the same as for Fig.~\ref{fig:obs250_unscaled_clusters_km}.}
                        \label{fig:nw012_unscaled_undeg_clusters_km}
\end{figure*}

\begin{figure*}
        \centering \hspace*{-5mm}
                \includegraphics[width=19.5cm]{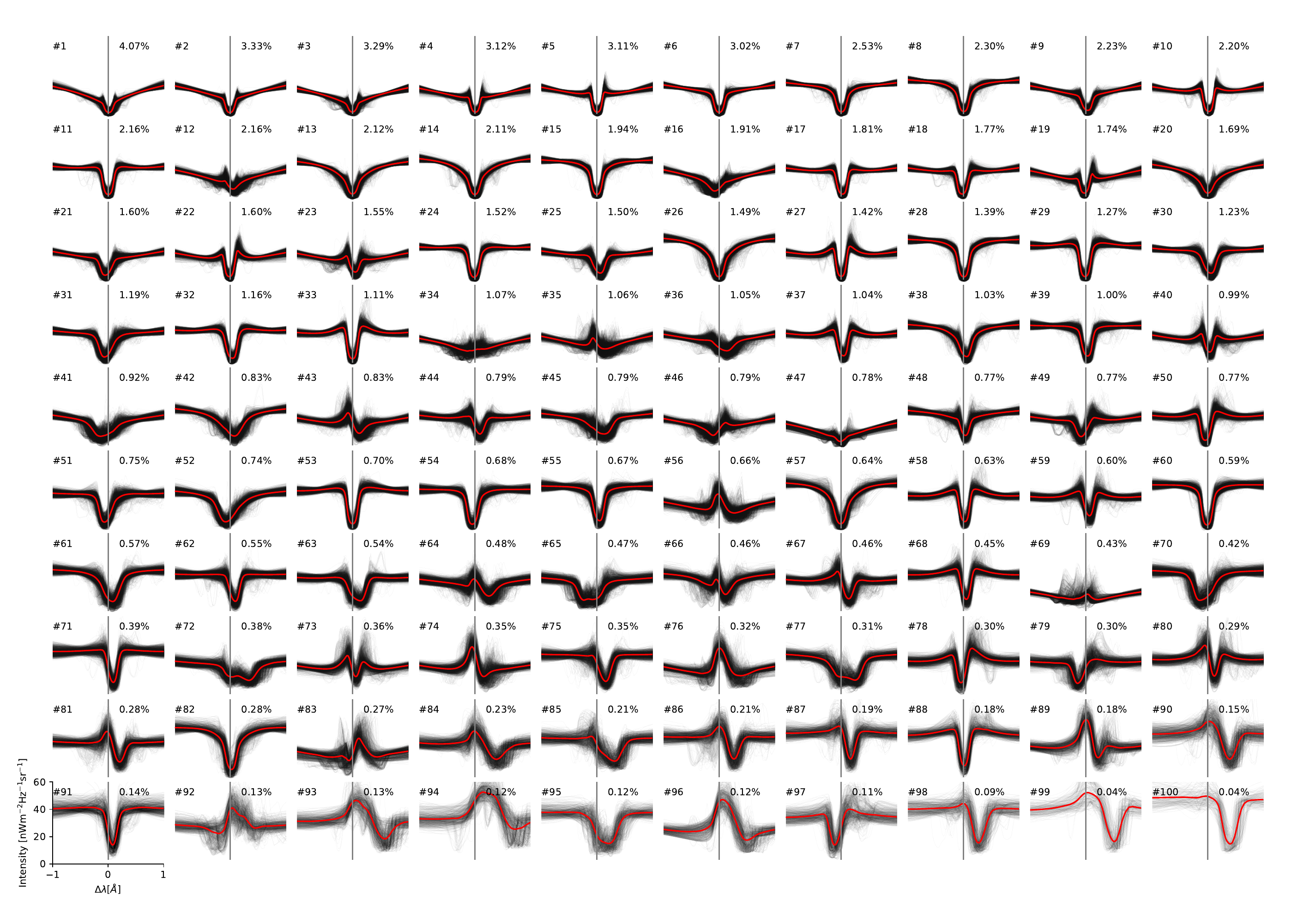}

                        \caption{$k$-means clusters of \ion{Ca}{II} 854.2~nm synthetic spectra from the \textit{nw072100} simulation. The legend is the same as for Fig.~\ref{fig:obs250_unscaled_clusters_km}.}
                        \label{fig:nw072_unscaled_undeg_clusters_km}
\end{figure*}

We employ the $k$-means and $k$-Shape \citep{10.1145/2723372.2737793} clustering methods on both synthetic and observed intensity profiles for the \ion{Ca}{ii} 854.2 nm line core. A thorough description for how these methods work, and how their results compare, can be found in \citetalias{2023A&A...675A.130M}. In short, they both iteratively partition a set of profiles amongst a predefined number $k$ clusters, grouping the profiles together based on some metric of similarity. For $k$-means that metric is the Euclidean distance, while $k$-Shape uses a more shape-based distance measure and also compares the profiles for a range of relative wavelength shifts. The $k$-Shape method assumes $z$-normalization, i.e. that each profile is scaled to have zero mean and unity variance. In practical terms, $k$-Shape is independent of the profiles' amplitudes and largely independent of the profiles' Doppler shifts, and it does, at least in some cases, do better at distinguishing profile shapes than k-means. It is, however, considerably slower computationally, and the amplitude invariance can group together profiles of rather different intensities. We use it here as a complementary tool to k-means, to check whether it generates clusters containing profile prototypes not seen in our k-means experiments.

We use the same libraries (scikit-learn, \citealt{scikit-learn}; tslearn, \citealt{JMLR:v21:20-091}) and methods ($k$-Means++ initialization, \citealt{10.5555/1283383.1283494} for the $k$-means method; some simple modifications of the tslearn library to make $k$-Shape run in parallel) as before. 

We perform the clustering not on the full line profile, but only in the central part within  $\lambda_0 \pm 0.1 ~\mathrm{nm}$, where $\lambda_0$ is the central wavelength of the \ion{Ca}{ii} 854.2 nm line. We made this choice for two reasons: this central region is the part formed in the chromosphere (already at 0.1~nm the line probes reversed granulation), and because our observations were limited to $\lambda_0 \pm 0.12 ~\mathrm{nm}$. In the rest of this work, when we refer to continuum we mean the local continuum at 0.1~nm, not the real continuum in the far wings. In order to give equal weight to all parts of the line profile, we interpolate our synthetic spectra to an equidistant grid of wavelength points in the range $\lambda_0 \pm 0.1 ~\mathrm{nm}$. The degraded synthetic spectra and the observations are given on the same equidistant grid of 21 wavelength points in the same range.

\section{Results}
\label{sec:results}

\subsection{Overview}
\label{sec:overview}

We have used the $k$-means clustering method with $k=100$ clusters and 10 re-initializations for the Stokes I profiles belonging to one snapshot for each of our simulations (to both degraded and non-degraded spectra), and to the observations. This particular choice of $k$ was made after experimentation as a trade-off between accuracy in the clustering, and human readability of the results. All cluster results shown in this manuscript stem from performing the clustering on the Stokes $I$ profiles without any normalization. For the synthetic profiles, the intensity units were \mbox{nW~m$^{-2}$~Hz~sr$^{-1}$}, and for the observations the intensities were not absolutely calibrated, so we used the arbitrary data number (DN) from the reduction. Additionally, we performed also both $k$-means and $k$-Shape clustering on the $z$-normalized intensity profiles to test whether that reveals any clusters with shapes not seen in the non-normalized spectra.

In an effort to provide some quantitative measures of the profile shapes in the following discussion, we define the depth of a profile as the difference between the maximum and minimum intensity in our considered wavelength window of $\lambda_0 \pm 0.1 ~\mathrm{nm}$. We also quantify the line widths by defining the full-width-half-maximum (FWHM) as the width between the points having intensities half-way between the minimum and maximum intensity in this wavelength window. We use these measures only on the mean profiles in clusters showing simple behavior, since they do not work well for more complicated line shapes.

\subsection{Spatially-averaged profiles}
\label{sec:mean_spectra}

Before moving to the clustering analysis we look at the spatially-averaged spectra. In Fig.~\ref{fig:mean_spectra} we plot the mean spectra, plus the \mbox{1-$\sigma$} variations around the mean for the observations and the three simulations. To allow a direct comparison, the synthetic spectra were degraded to the observational conditions before computing the mean and 1-$\sigma$ variations, and all spectra were normalized by the local continuum at $\lambda_0 + 0.1$~nm.

A noteworthy difference is that the simulations, even after spatial and spectral degradation, have spatial variations that are about twice as large as the variations in the observations. And the amount of variation does not seem to change much from the more quiet \textit{ch012023} to the more active \textit{nw072100}. The \textit{nw012023} mean profile is redshifted because the particular snapshot we used has a net downflowing atmosphere, leading to a shifted and more asymmetric mean profile. In terms of line width, the observations are broader than all simulations, but both \textit{ch012023} and \textit{nw012023}, which have a horizontal grid size of 23~km, are much closer to the observations than \textit{nw072100}, which has a grid size of 100~km. In numbers, the FWHM of these mean profiles are 66 pm, 55 pm, 49 pm, and 31 pm, respectively, for the observations, \textit{ch012023}, \textit{nw012023}, and \textit{nw072100}. Without spectral and spatial degradation, we obtained FWHMs of 53 pm, 48 pm and 25 pm, respectively, for the mean profiles from \textit{ch012023}, \textit{nw012023}, and \textit{nw072100}. We note that when we discuss the FWHMs of the synthetic profiles in the following sections, we refer to the the undegraded profiles at native spatial and spectral resolution.

\subsection{Stokes $I$ clusters}
\label{sec:clusters}

\subsubsection{Observations}
\label{sec:obs_clusters}

We show the resulting clustering for our observations in Fig.~\ref{fig:obs250_unscaled_clusters_km}. This clustering was performed for a single CRISP scan (about $10^6$ spectra in total). We also experimented with scans taken at different times, and clustering multiple scans at the same time, but find little variation in the results. The most frequent types of clusters hardly change, and the few differences are mostly in the least frequent clusters, which can vary slightly from scan to scan. Hereafter, we discuss only the observations shown in Fig.~\ref{fig:obs250_unscaled_clusters_km}, since they were taken with some of the best conditions and we find them representative of the general properties of the observed region.

For the most part the observed spectra appear quite tightly constrained, with little variation inside most clusters, and typical line shapes not too dissimilar between clusters. The majority of clusters present absorption profiles with fairly wide line cores and gently curving transitions from the inner wings to the core; prime examples are e.g. \#10 or \#44. Most of the variation for these profile types comes as gentle Doppler shifts of the line, sometimes accompanied by a slight asymmetry (e.g. \#46 or \#87), or a larger asymmetry (e.g. \#85). There is also some variation in the width of the profiles, and how steep the transitions from the wings to the core are; compare, for instance, \#30 with \#91. Additionally, there is some variation in local continuum and line depth; contrast, for example, \#98 with \#62. On the whole, however, the general shapes are similar.

On the other hand, there are some `families' of clusters that break the mold. One distinct type is the very shallow, sometimes almost triangular, profiles of clusters like \#45, \#60, \#66, \#81, \#84, \#63 or  \#90,. Some of these clusters include profiles similar to the `raised core' profiles found by \citet{2013ApJ...764L..11D}, although the magnetic configuration of our observations is somewhat different from those of \citet{2013ApJ...764L..11D}, with a smaller magnetic canopy. There is also cluster \#99 which displays emission on the left side of the core, similar to the chromospheric bright grain-like (CBG-like) profiles we studied in \citetalias{2023A&A...675A.130M}. This is the only cluster which shows very clear emission, but there are a few cases (\#77, \#79, \#80, \#88, \#89) where there are some cluster members showing enhanced intensities around the `elbow' marking the transition from wings to core. Beyond those, there are the three clusters \#93, \#97, and particularly \#100,  which are somewhat less constrained than the others, and display more complex shapes. 

As mentioned previously, we have also used the $k$-Shape method alongside $k$-means to cluster these profiles after applying $z$-normalization. The purpose of this is to ensure that we get a more complete view of which profile shapes are present in our data. Those experiments yielded qualitatively very similar results as for the unnormalized case, with the largest difference being that they picked up and separated out a few clusters showing flat-bottomed or complex line cores which mostly belong to clusters such as  \#63, \#81, \#90, \#97, \#100, in Fig. \ref{fig:obs250_unscaled_clusters_km}. That is not surprising, as the $z$-normalization amplifies the relative differences in amplitude between the members of the shallow clusters, making their shapes more distinct. As for the profiles in \#100, z-normalization reduced the difference in absolute intensity between members of that cluster and the other clusters, so that the different shapes present in that cluster could be separated and put into other clusters more based on the shape than the amplitude. For our purposes in this paper, we are interested in both the profiles shapes and their absolute amplitudes, so we focus on the clusters with unnormalized profiles; however, we would like to emphasize that there is some diversity in the shapes found for the least constrained clusters.

\subsubsection{\textit{ch012023} clusters}
\label{sec:ch012023_clusters}

We now turn to the clusters retrieved from our synthetic observations. We carried out the clustering for the original and degraded synthetic spectra. We find the same general trends for the retrieved clusters in both cases, with the main difference being reduced variations in each cluster, and a reduced range in the continuum variations (as expected from the spatial degradation) in the clusters of degraded spectra. Since we later look into the atmospheric structures for some of the synthetic clusters, we focus our analysis on the undegraded profiles, as the spatial and spectral degradation makes it difficult to assign unequivocal values of the atmospheric parameters to the degraded spectra. We show the \textit{ch012023} clustering results for the original resolution in Fig.~\ref{fig:ch012_unscaled_undeg_clusters_km}, while a similar figure for the degraded cases is shown in Appendix \ref{sec:appendix_degraded_clusters}.

The \textit{ch012023} simulation represents a quiet Sun scene, with magnetic fields resembling the conditions of a coronal hole. Thus it is noteworthy that in Fig.~\ref{fig:ch012_unscaled_undeg_clusters_km} so many clusters show profiles with emission features and strong asymmetries. This is an important difference from the clusters found in the observations, which contain mostly absorption profiles. Among the \textit{ch012023} clusters we find CBG-like profiles (for instance \#61, \#70, \#71, \#89, \#93, \#94) and the double-peaked profiles (seen in \#41, \#81, \#99) discussed in \citetalias{2023A&A...675A.130M}, to the complicated and poorly constrained clusters such as \#30, \#36, \#60. Furthermore, cluster \#100 displays reversed CBG-like profiles (namely, the emission is on the red side of the core) which are not clearly present in the observations. Beyond that, we have cases of enhanced `elbows' (e.g. \#63, \#77, \#78), where there is emission around either the blue, the red, or both transitions from wings to core; these features are only weakly seen in the observed clusters. We also find sharply asymmetric absorption profiles (for instance \#22, \#87, \#95,) and flat-bottomed profiles (e.g. \#45, \#67) that do not match the clusters in the observations.

Not only are the shapes found more varied and the number of clusters with exotic shapes larger, but the number of profiles belonging to these atypical clusters is far larger in the simulation than in the observations. As an example, just the three clusters \#61, \#70, and \#71 in Fig. \ref{fig:ch012_unscaled_undeg_clusters_km} contain a larger percentage of profiles  (roughly 1.7 \% vs. 1.4 \% of all profiles), than all the observed clusters with clear emission features (\#77, \#79, \#80, \#88, \#89, \#99 in Fig. \ref{fig:obs250_unscaled_clusters_km}). 

There are, of course, profiles that appear similar between the observations and simulation as well. We find several more typical absorption profiles, for instance \#23, \#34, \#51, \#80, \#84; and some of the more shallow profiles, such as \#4 and \#8 in Fig. \ref{fig:ch012_unscaled_undeg_clusters_km} bear strong resemblance to their observed counterparts, such as \#63 and \#81 in Fig. \ref{fig:obs250_unscaled_clusters_km}. However, there are some marked differences between the typical absorption profiles found in the observations versus the simulation. The most noticeable difference is the width of the profiles, as the synthetic profiles are narrower the observed profiles. There are, however, variations in how large this difference is; the wider profiles in Fig. \ref{fig:ch012_unscaled_undeg_clusters_km}, such as \#34 and \#51 (FWHM of 47 pm and 40 pm, respectively), are not that much narrower than \#58, \#69, \#83, or \#91 in Fig. \ref{fig:obs250_unscaled_clusters_km} (FWHMs of 50 pm, 50 pm, 48 pm, and 49 pm, respectively) . On the other hand, the narrow synthetic profiles, such as \#32, \#57, \#69, or \#75 in Fig. \ref{fig:ch012_unscaled_undeg_clusters_km} (FWHMs of 29 pm, 25 pm, 27 pm, and 26 pm, respectively), contrast greatly with the observational clusters. 

Another difference is the shape of the transition from wings to core; the synthetic profiles generally have a much steeper, in cases almost cliff-like, transition than the observed profiles, which tend to exhibit far smoother and more gently curving slopes. This difference is particularly noticeable for narrow profiles such as \#69, but it also occurs for wider profiles like \#84. More similar to the observations, in terms of shape, are the profile clusters like \#9, \#14 and \#23 in Fig. \ref{fig:ch012_unscaled_undeg_clusters_km}, which compare pretty well with e.g. \#43, \#48, \#57, \#77 in the observation clusters shown in Fig. \ref{fig:obs250_unscaled_clusters_km}. In sum, we find large differences between the clusters obtained from our observation and from our quietest simulation, though there are some instances were there are strong resemblances between them.

As with the observed spectra, we have also done $k$-Shape and $k$-means clustering using $z$-normalization on both these and the other synthetic spectra. Similarly to the observational case, but even more pronounced, we find several clusters with rather complicated line core shapes, which correspond to the least constrained or most shallow clusters for the unnormalized intensities. These complicated line core shapes include flat-bottomed cores, W-shaped central reversals, and M-shaped double-peaked central reversals that both do and do not exceed the intensities of the nearby inner wings. Again, a longer discussion about the varieties of shapes revealed when ignoring the absolute intensities is beyond the scope of the current work, but we wish to emphasize that there are complicated spectral profiles found both in the least constrained clusters like \#30, \#55, \#73, \#99, and also in the seemingly simple shallow shapes like \#3, \#4 \#11, \#18 in Fig. \ref{fig:ch012_unscaled_undeg_clusters_km}. In appendix \ref{sec:appendix_degraded_clusters}, we briefly discuss and show the clustering results obtained with $k$-Shape on the $z$-normalized profiles for the observation and the \textit{ch012023} simulaiton. While we have not studied the $z$-normalized clustering results for the other simulations' synthetic spectra in detail, at first impression they seem to display the same tendencies as in this case.

\subsubsection{\textit{nw012023} clusters}
\label{sec:nw012023_clusters}

In Fig. \ref{fig:nw012_unscaled_undeg_clusters_km} we show the results of applying the same type of clustering to the more magnetically active simulation \textit{nw012023}. On the whole the results are quite similar to what we previously saw for the quiet \textit{ch012023} simulation, with all the same shapes found in one being mirrored in the other. The primary difference in the clustering results for the \textit{nw012023} simulation compared to the \textit{ch012023} simulation is that the intensity values span a wider range; that is, some profiles are deeper, and emission peaks can also reach higher. Some of the clusters also seem to display more variance, meaning the cluster centroids seem to be less representative of the individual cluster members and there is likely quite a bit of mixing between different shapes in those clusters, e.g. \#9,  \#36, \#61, \#74 in Fig. \ref{fig:nw012_unscaled_undeg_clusters_km}. Similar poorly constrained clusters were also seen for the \textit{ch012023} simulation, e.g. \#73 and \#93 in Fig. \ref{fig:ch012_unscaled_undeg_clusters_km}, but here they seem to be even stronger. 

As can be seen in appendix \ref{sec:appendix_degraded_clusters}, the effect of spatial and spectral degradation is a major decrease in within-cluster variance, but the key differences to the observation clusters remain.

\subsubsection{\textit{nw072100} clusters}
\label{sec:nw072100_clusters}

Finally, we show the clustering results for the \textit{nw072100} simulation in Fig. \ref{fig:nw072_unscaled_undeg_clusters_km}. In this case we still see more extreme shapes than in the other two simulations, but many of the cluster shapes are similar. The variance within clusters appears to be quite a bit larger than before (e.g. \#24,\#69, \#92). The profile amplitudes are also generally larger; that is to say, many of the emission features have higher peak intensities and several of the absorption profiles are deeper. This is as expected since this simulation is more active and vigorous than the other two. There is one new `family' of clusters that appears for this simulation, namely the very broad emission profiles in clusters \#93, \#94, and \#96. These profiles appear to have somewhat similar shapes, but much broader, to the CBG-like profiles seen in both the previous cluster results and in other clusters for this simulation (e.g. \#74, \#84, \#89). However, looking into their atmospheric structure (plotted in Fig. \ref{fig:nw072_broad_emission_41_79_48_96}) we find that \#93, \#94, and \#96 are associated with high temperatures ( $\gtrsim 7$ kK) and fast downflows ($\ll -8$ km s$^{-1}$)  in the range $ -6 \leq \log\tau_{5000} \leq -4$, where $\tau_{5000}$ is the optical depth for light at 500 nm (5000 Å), and show strong ($> 10$ mT in absolute value) vertical magnetic field components over most of the line-forming region. 

In terms of the profile widths for the more typical absorption profiles, they are narrower in this simulation compared to the other two simulations; the narrowest profile clusters in Fig. \ref{fig:nw072_unscaled_undeg_clusters_km} (e.g. \#55, \#71, \#88, with, respectively, FWHMs of 22 pm, 19 pm, 19 pm) are clearly less wide than any of the clusters we see in Fig. \ref{fig:ch012_unscaled_undeg_clusters_km} or Fig. \ref{fig:nw012_unscaled_undeg_clusters_km}. However, there are also several clusters showing quite wide absorption profiles (e.g. \#52, \#61, \#95, in Fig. \ref{fig:nw072_unscaled_undeg_clusters_km}, with FWHMs of, respectively, 42 pm, 36 pm, 44 pm), that seem comparable with the moderately wide clusters for the two higher-resolution simulations. That the typical absorption profiles are narrower in \textit{nw072100}, which has the lowest spatial resolution, is consistent with the findings of \citet{2023ApJ...944..131H}.

\subsection{Atmospheric structure for selected clusters}
\label{sec:atmos_structure}

\begin{figure*}
        \centering \hspace*{-5mm}
                \includegraphics[width=19.5cm]{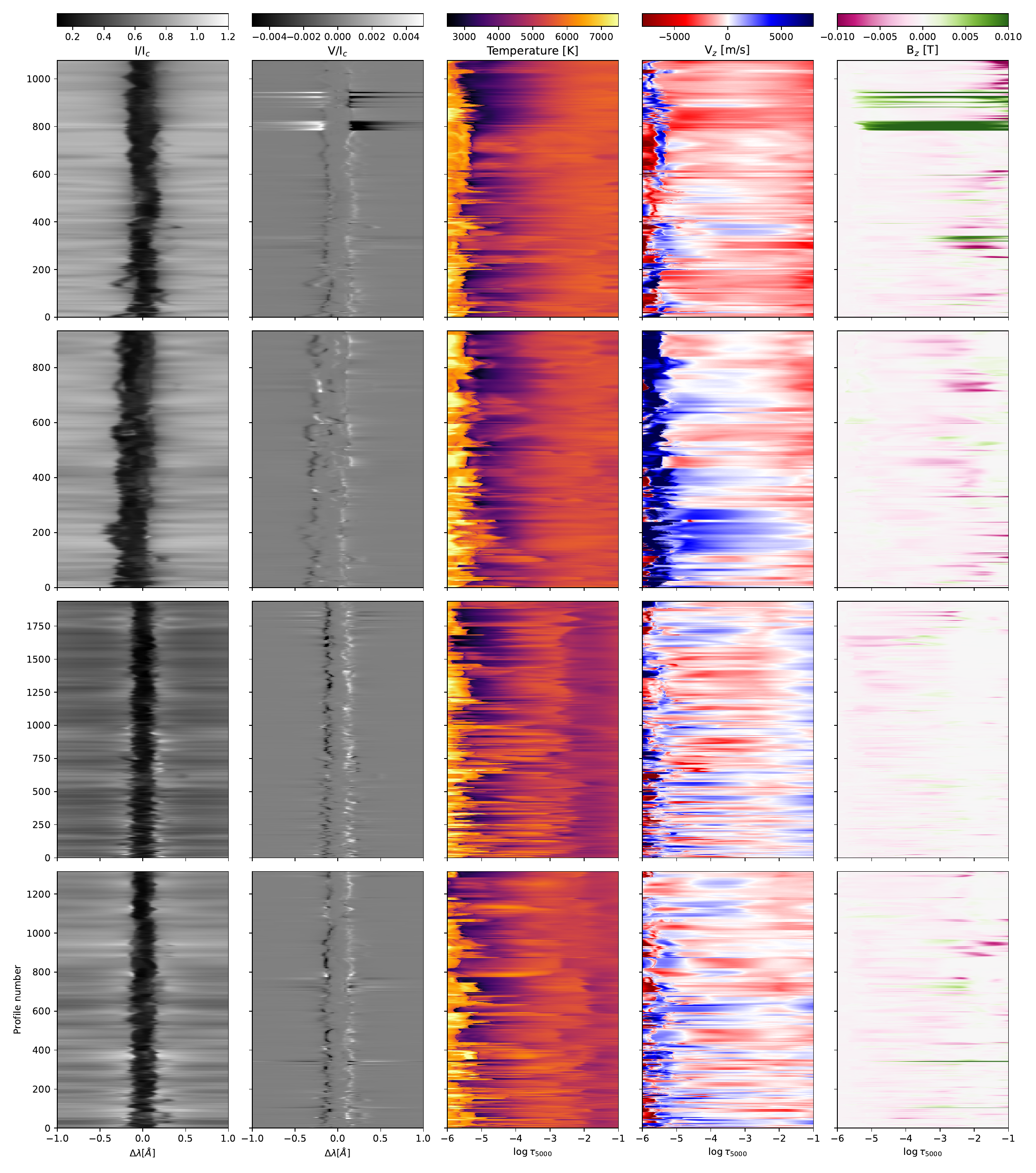}

                        \caption{Individual spectra and atmospheric structure of selected clusters from the \textit{ch012023} simulation. From left to right, the first two columns show the Stokes $I$ and $V$ (normalized to the nearby continuum, around $\lambda_0 + 0.95$ nm), while the last three columns show respectively the temperature, vertical velocity, and vertical magnetic field as a function of $\log\tau_{5000}$. Each row depicts a different cluster. From the top, the first and second are respectively \#80 and \#84 (from the numbering in Fig. \ref{fig:ch012_unscaled_undeg_clusters_km}), and represent some of the wider line profiles. The third and fourth rows depict cluster numbers \#57 and \#75, which represent some of the narrowest line profiles.}
                        \label{fig:ch012_example_31_41_70_56}
\end{figure*}

\begin{figure}
        \centering
                \includegraphics[width=0.49\textwidth]{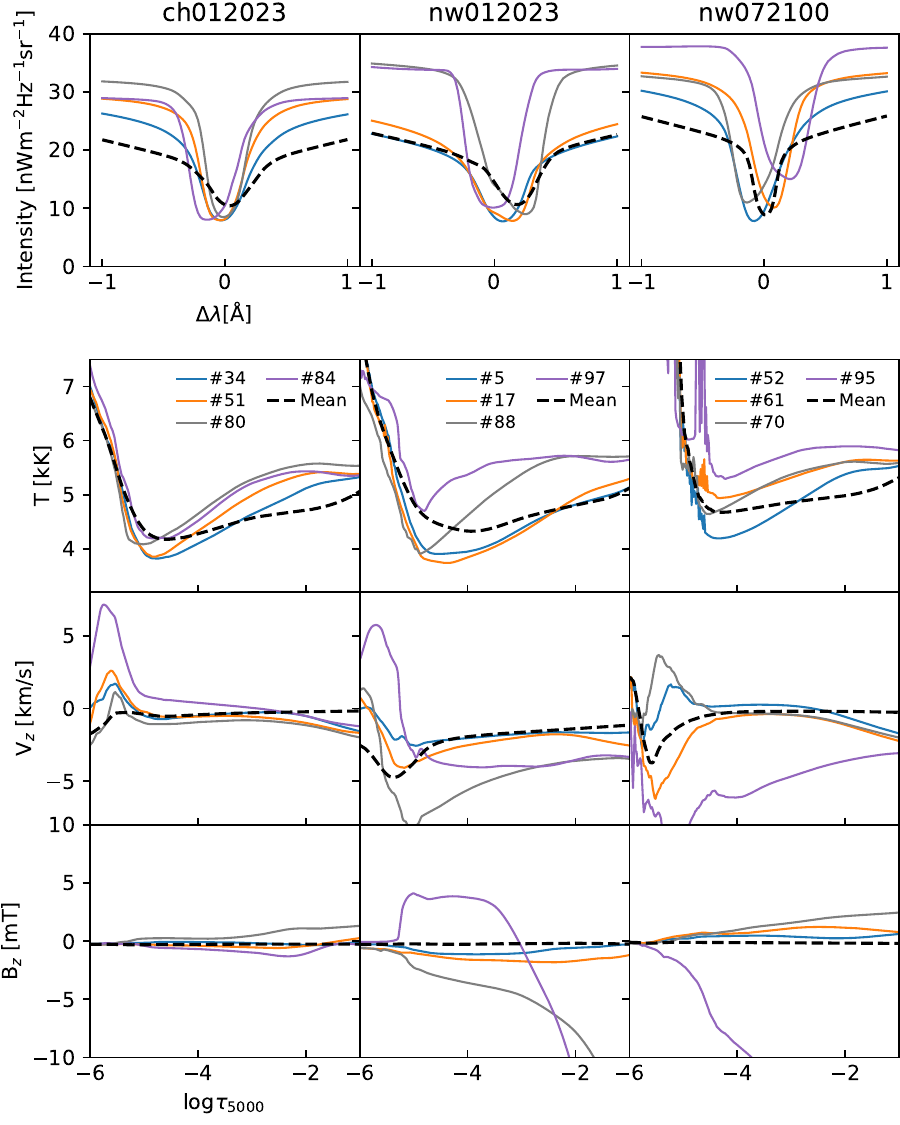}

                        \caption{Mean spectra and atmospheric structure for selected clusters representing wide line profiles. Each column depicts four selected clusters for each of the three simulations, in addition to the mean for the full simulation box (dashed black line). The top row shows the mean spectra for each of the clusters, while the bottom three rows show the temperature, vertical velocity, and vertical magnetic field as a function of $\log\tau_{5000}$. The cluster numbers for each simulation are indicated in the legend of the temperature plot.}
                        \label{fig:wide_atmos_means}
\end{figure}

\begin{figure}
        \centering
                \includegraphics[width=0.49\textwidth]{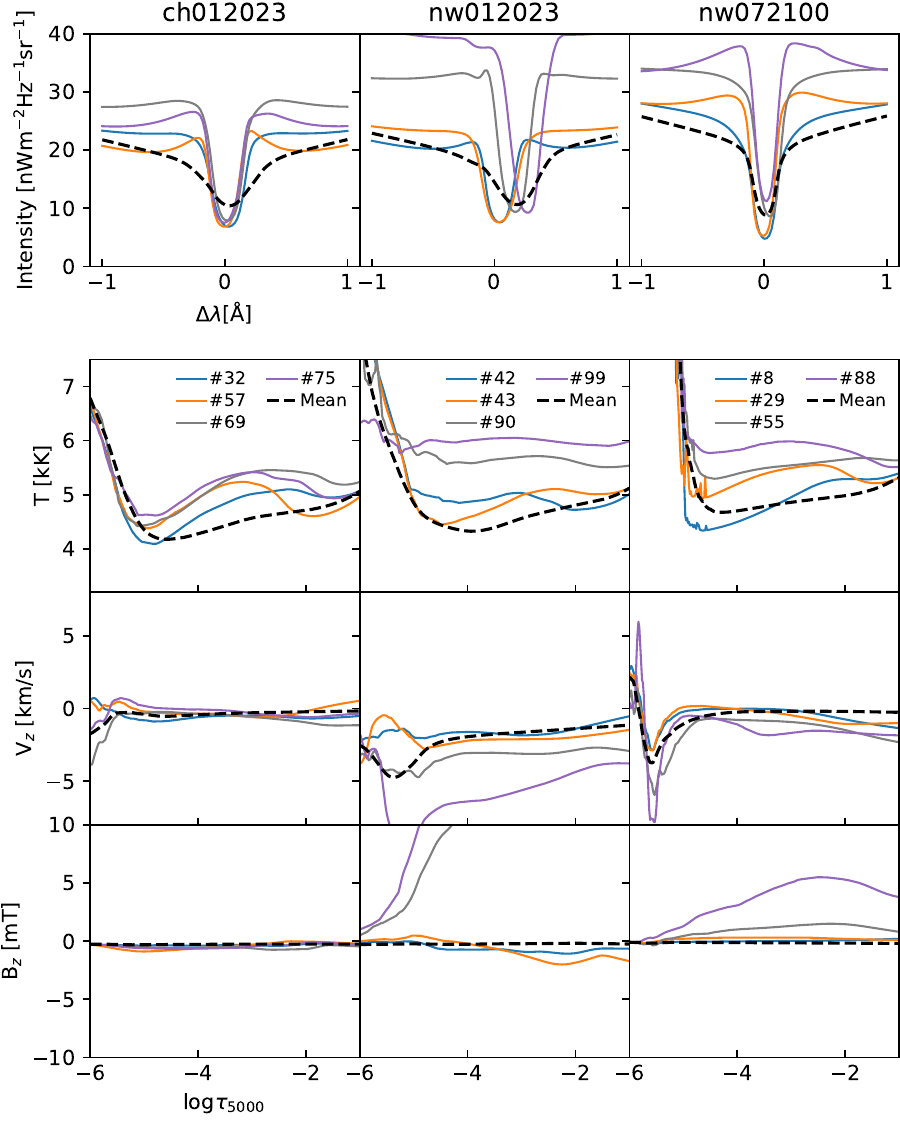}

                        \caption{Mean spectra and atmospheric structure for selected clusters representing narrow line profiles. The legend is the same as for Fig.~\ref{fig:wide_atmos_means}.}
                        \label{fig:narrow_atmos_means}
\end{figure}

\begin{figure}
        \centering
                \includegraphics[width=0.49\textwidth]{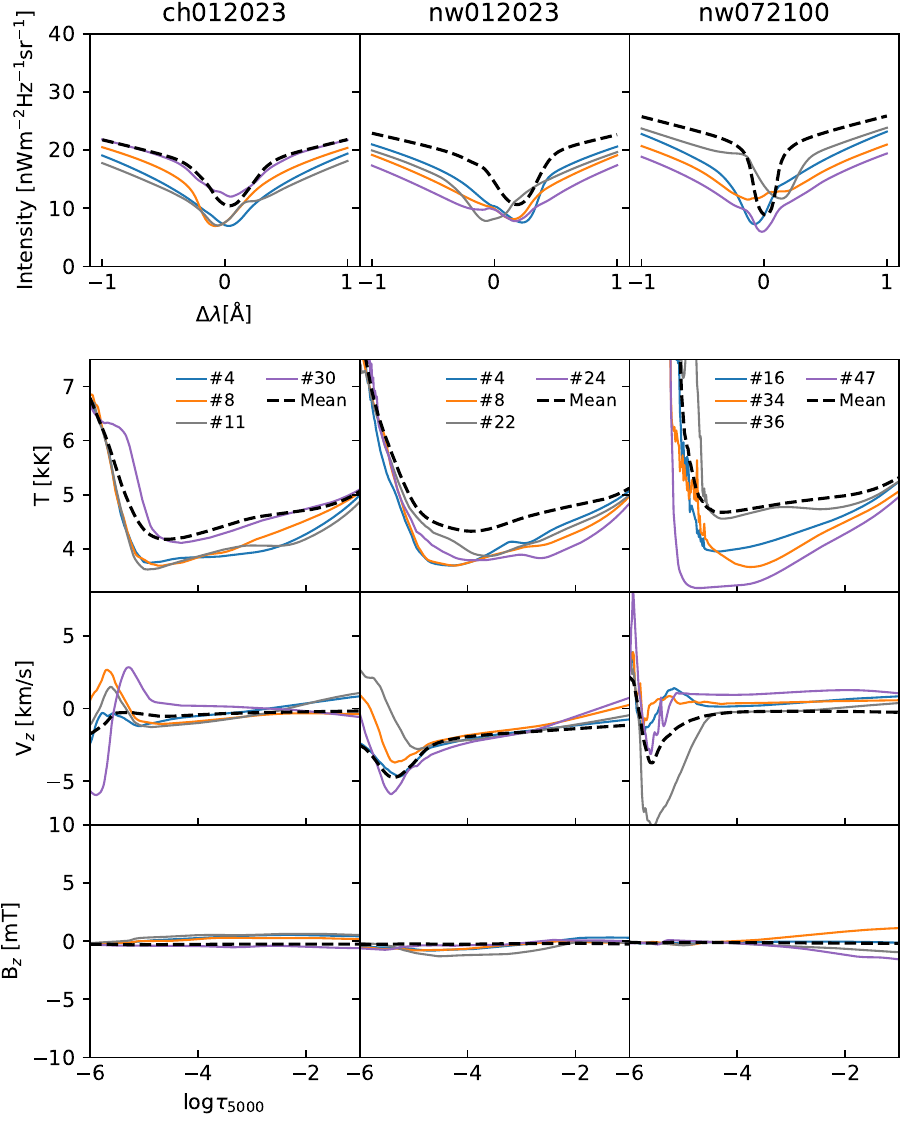}

                        \caption{Mean spectra and atmospheric structure for selected clusters representing shallow line profiles. The legend is the same as for Fig.~\ref{fig:wide_atmos_means}.}
                        \label{fig:shallow_atmos_means}
\end{figure}

\begin{figure}
        \centering
                \includegraphics[width=0.49\textwidth]{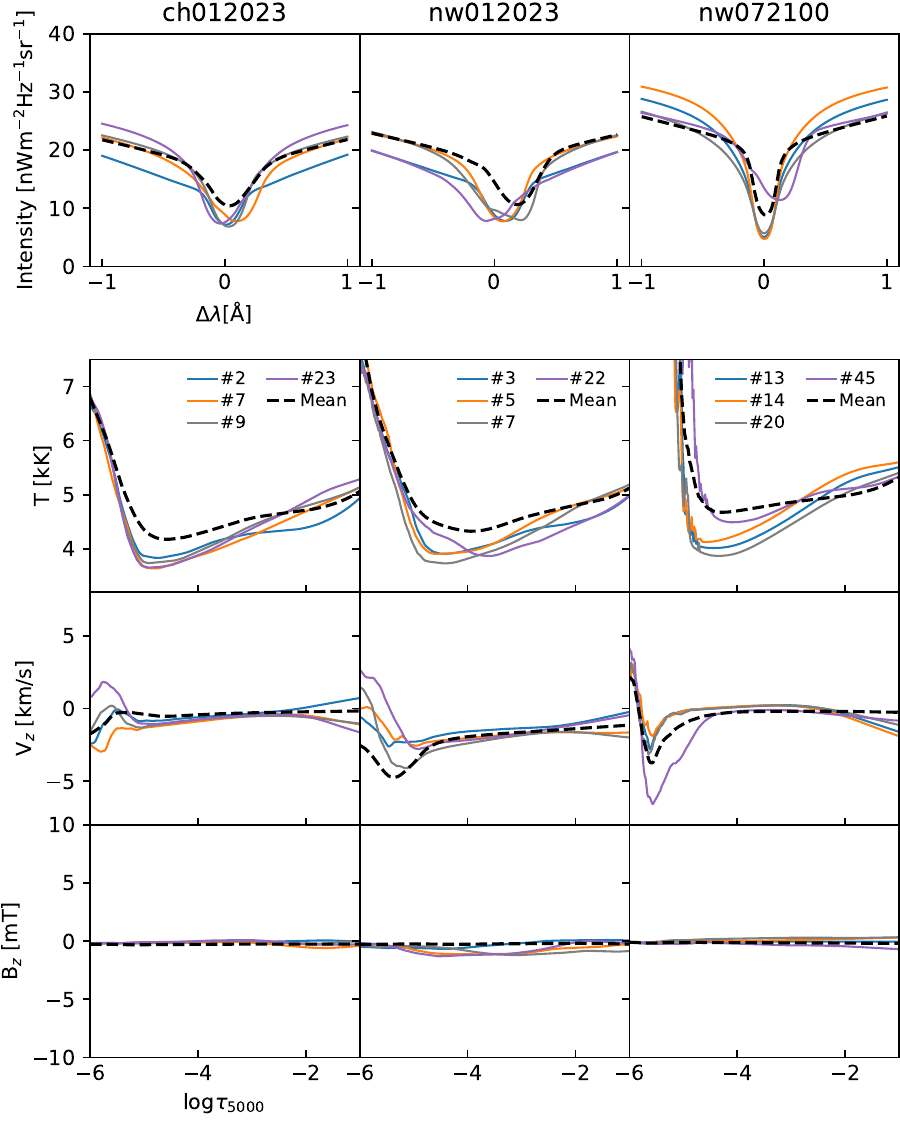}

                        \caption{Mean spectra and atmospheric structure for selected clusters representing the line profiles with the smoothest transition from wing to core. The legend is the same as for Fig.~\ref{fig:wide_atmos_means}.}
                        \label{fig:leastcurved_atmos_means}
\end{figure}

So far the clustering results have shown us how similar, and dissimilar, the profiles are amongst the simulations and compared to the observation. We now take a detailed look at the atmospheric structure for a few selected clusters of synthetic profiles, in order to see which trends in the atmospheric parameters correlate with these cluster `families'. We focus on four `families` of clusters, all whose mean profiles are in absorption. The first is composed of wide profiles: profile shapes that show a broad line core, and typically stronger continuum intensities. The second is composed of narrow profiles: clusters that have some of the narrowest line widths. The third is composed of shallow profiles: lines with the smallest difference between wing and core intensity. Finally, the fourth family is composed of clusters of line profiles that have the gentlest (i.e. most gradual) transition from line wings to line core. For each of these families, we selected four representative clusters from each of the simulations. 

We elect to look at the widest and the narrowest profiles because the line width is one of the more obvious discrepancies between the simulations and observations, and it is therefore interesting to see which type of atmospheric structures can gives rise to narrow and broad lines. This might in turn provide indications to what our simulations are lacking compared to the real sun. As for the other two families, the shallow and the gently curving clusters, we chose to investigate them because they are very reminiscent of clusters from the observations (e.g. cluster \#81 in Fig. \ref{fig:obs250_unscaled_clusters_km} appears quite similar to \#4 in Fig. \ref{fig:ch012_unscaled_undeg_clusters_km}), and thus their atmospheric structures likely resemble more closely conditions in parts of the solar atmosphere.

The clusters for each family were manually selected, as it is somewhat difficult to formulate quantitative criteria for whether a cluster belongs in one `family' or another. As such, there is some subjectivity involved in our particular choice of clusters for detailed analysis. That is why we look at several examples for each cluster type and simulation, as we intend to discover the qualitative trends in atmospheric structure for the cluster `families' without relying on single examples.

Furthermore, it is important to recognise that there can be quite wide variations within clusters, and that even though the cluster means may appear well behaved, they are not necessarily fully representative of the individual profiles that make up the clusters. Therefore, the way we have investigated these clusters is by carefully examining plots like Fig. \ref{fig:ch012_example_31_41_70_56}, which displays the atmospheric structure for each profile in the clusters \#80, \#84, \#57, \#75 from  Fig. \ref{fig:ch012_unscaled_undeg_clusters_km}, which are, respectively, two of the wider and two of the narrower clusters found for \textit{ch012023}. The individual profiles making up the clusters are stacked along the vertical axis, whose number just means profile number. The leftmost columns show the Stokes $I$ and Stokes $V$ profiles, normalized to the nearby continuum at approximately $\lambda_0 + 0.95$ nm, as a function of wavelength. The three rightmost columns show the stratification of temperature, line-of-sight velocity, and line-of-sight magnetic field strength against $\log\tau_{5000}$. These plots show the individual variations within the clusters, and make clearer the less common atmospheric features which are not as easily seen when considering averages. As an example of this, some of the narrow profiles have enhanced temperatures stretching across the whole range $ -5 <\log\tau_{5000} < -2 $ that correlate with emission in the transition from line wings to line core. For the sake of brevity, we show in Fig.~\ref{fig:ch012_example_31_41_70_56} only a few representative clusters in detail. For the remaining selected clusters in each group, we show only a summarized view as in Fig. \ref{fig:wide_atmos_means}. However, we analyzed each of the selected clusters in detail. To aid in the comparison of the clusters across both type and family, we provide an overview of the estimated line widths and depths for mean profiles of selected clusters in Table. \ref{tab:rp_quantities_ch012}, Table. \ref{tab:rp_quantities_nw012}, and Table. \ref{tab:rp_quantities_nw072}

\begin{table}
\caption{Line FWHMs and depths for selected clusters from simulation \textit{ch012023}}
    \centering
    \begin{tabular}{cccc}
        \hline\hline \rule{0pt}{2.ex}Profile type & Cluster & FWHM  & Line depth  \\
           & & (pm) & ($\mathrm{nW m^{-2}}$ $ \mathrm{Hz^{-1} sr^{-1}}$) \\
        \hline
        Wide & \#34 & 47 & 18.4 \\
        Wide & \#51 & 40 & 20.9 \\
        Wide & \#80 & 36 & 23.3 \\
        Wide & \#84 & 43 & 20.9 \\
        \hline Narrow & \#32 & 29 & 16.5 \\
        Narrow & \#57 & 25 & 16.5 \\
        Narrow & \#69 & 27 & 20.7 \\
        Narrow & \#75 & 26 & 19.1 \\
        \hline Shallow & \#4 & 81 & 12.4 \\
        Shallow & \#8 & 59 & 13.5 \\
        Shallow & \#11 & 83 & 11.1 \\
        Shallow & \#30 & 69 & 9.82\\
        \hline Smooth transition & \#2 & 49 & 12.0 \\
        Smooth transition & \#7 & 58 & 14.3 \\
        Smooth transition & \#9 & 42 & 15.7 \\
        Smooth transition & \#23 & 54 & 17.2 \\
        \hline
    \end{tabular}
    \label{tab:rp_quantities_ch012}
\end{table}

\begin{table}
\caption{Line FWHMs and depths for selected clusters from simulation \textit{nw012023}}
    \centering
    \begin{tabular}{cccc}
        \hline\hline \rule{0pt}{2.ex}Profile type & Cluster & FWHM  & Line depth  \\
           & & (pm) & ($\mathrm{nW m^{-2}}$ $ \mathrm{Hz^{-1} sr^{-1}}$) \\
        \hline
        Wide & \#5 & 47 & 15.1 \\
        Wide & \#17 & 56 & 17.2 \\
        Wide & \#88 & 53 & 25.9 \\
        Wide & \#97 & 46 & 24.2 \\
        \hline Narrow & \#42 & 24 & 14.2 \\
        Narrow & \#43 & 31 & 16.5 \\
        Narrow & \#90 & 26 & 24.3 \\
        Narrow & \#99 & 28 & 31.5 \\
        \hline Shallow & \#4 & 66 & 13.4 \\
        Shallow & \#8 & 84 & 11.1 \\
        Shallow & \#22 & 69 & 12.1 \\
        Shallow & \#24 & 100 & 9.6 \\
        \hline Smooth transition & \#3 & 47 & 12.1 \\
        Smooth transition & \#5 & 47 & 15.1 \\
        Smooth transition & \#7 & 63 & 15.2 \\
        Smooth transition & \#22 & 69 & 12.1 \\
        \hline
    \end{tabular}
    \label{tab:rp_quantities_nw012}
\end{table}

\begin{table}
\caption{Line FWHMs and depths for selected clusters from simulation \textit{nw072100}}
    \centering
    \begin{tabular}{cccc}
        \hline\hline \rule{0pt}{2.ex}Profile type & Cluster & FWHM  & Line depth  \\
           & & (pm) & ($\mathrm{nW m^{-2}}$ $ \mathrm{Hz^{-1} sr^{-1}}$) \\
        \hline
        Wide & \#52 & 43 & 22.4 \\
        Wide & \#61 & 36 & 23.2 \\
        Wide & \#70 & 40 & 21.7 \\
        Wide & \#95 & 44 & 22.8 \\
        \hline Narrow & \#8 & 24 & 23.2 \\
        Narrow & \#29 & 21 & 24.6 \\
        Narrow & \#55 & 22 & 25.4 \\
        Narrow & \#88 & 19 & 27.2 \\
        \hline Shallow & \#16 & 56 & 15.9 \\
        Shallow & \#34 & 98 & 9.5 \\
        Shallow & \#36 & 44 & 12.2 \\
        Shallow & \#47 & 74 & 13.5 \\
        \hline Smooth transition & \#13 & 36 & 23.8 \\
        Smooth transition & \#14 & 31 & 26.2 \\
        Smooth transition & \#20 & 40 & 20.9 \\
        Smooth transition & \#45 & 48 & 14.9 \\
        \hline
    \end{tabular}
    
    \label{tab:rp_quantities_nw072}
\end{table}

\subsubsection{Wide profiles}
\label{sec:wide_profiles}

 In Fig. \ref{fig:wide_atmos_means} we show the spectra and atmospheric quantities averaged over four different clusters of each simulation, along with the mean spectrum and atmospheric quantities averaged over the full simulation box. The averaging of the atmospheric quantities was performed over $\tau_{5000}$ isosurfaces. The clusters we selected correspond in Fig. \ref{fig:wide_atmos_means} to clusters with wide line profiles, and their numbers are \#34,  \#51, \#80, \#84 from \textit{ch012023} (Fig. \ref{fig:ch012_unscaled_undeg_clusters_km}); clusters \#5, \#17, \#88, \#97 from \textit{nw012023} (Fig. \ref{fig:nw012_unscaled_undeg_clusters_km}); clusters \#52, \#61, \#70, \#95 from \textit{nw072100} (Fig. \ref{fig:nw072_unscaled_undeg_clusters_km}). 

For the \textit{ch012023} case there appears to be a common trend for the wider clusters in terms of temperature, vertical velocity and vertical magnetic field strength. The temperature goes from a moderately hot bottom up to a cold layer which extends to the end of the line forming region, where $\log\tau_{5000}$ is approximately between $-5.5$ and $-5$. The velocities are mostly weak or moderate (absolute values of $<2.5$ km s$^{-1}$), as are the vertical magnetic field strengths (absolute values of $< 5$ mT). A slight exception to the general tendencies are the profiles around profile number 200 of cluster \#84, seen in the second row of Fig. \ref{fig:ch012_example_31_41_70_56}. These show a noticeable widening on the blue side, and here the atmospheric structure shows a region of high temperatures extending below $\log\tau_{5000} = -5$ coinciding with a moderately strong upflow ($v_z \sim 3$ km s$^{-1}$). 

For the \textit{nw012023} case, we find that there are two general types of atmospheres that produce the wide clusters. The first type, \#5 and \#17 from Fig. \ref{fig:nw012_unscaled_undeg_clusters_km} go from an averagely warm bottom, via a fairly constant gradient, to extended cold layers around $\log\tau_{5000} \approx -5$ and below. The atmospheres of these profiles have weak to moderate vertical velocities of the downflowing variety ($v_z > -5$ km s$^{-1}$), along with weak to moderate magnetic field strengths of either polarity (absolute values of $< 5$ mT), throughout the line forming region. The other type, \#88 and \#97 from Fig. \ref{fig:nw012_unscaled_undeg_clusters_km}, start with hotter bottom layers where the temperature does not decrease much before $\log\tau_{5000} \approx -3$, or sometimes not even before $\log\tau_{5000})\approx -5$. These wide profiles with enhanced temperatures correlate with moderate-to-strong downflowing velocities ($v_z < -5$ km s$^{-1}$) reaching up to $\log\tau_{5000} \approx -5$, at which point strong upflows ($v_z > 5$ km s$^{-1}$) appear. They also coincide with strong magnetic fields of both polarities (absolute values of $>10$ mT), which do however taper off around $\log\tau_{5000} \approx -5$. The hotter lower atmospheres in these types of profiles also help explain why the local continuum is much higher, and since the temperature gradient is shallow until $\log\tau_{5000}\approx -3$, and only then gets steep, the transition from continuum to line core in the profile is more abrupt, in contrast with clusters \#5 and \#17, whose mean profiles have a smoother transition from wing to core.

For the \textit{nw072100} case, the trend across all four clusters is that they start at higher than average temperatures at the bottom, and decrease to a minimum around $\log\tau_{5000} \approx -4.5$  The vertical velocities are mostly weak ($< 2.5$ km s$^{-1}$) up to $\log\tau_{5000} \approx -5$, except for the hottest cluster with the shallowest temperature gradient (\#95 in Fig. \ref{fig:nw072_unscaled_undeg_clusters_km}), which has a lot of moderately strong downflows ($v_z \leq -2.5$ km s$^{-1}$) throughout the line forming region. This is also the cluster with the strongest vertical magnetic field strengths (absolute values of $> 10$ mT), although the other clusters also have some moderately strong fields present. In all clusters both magnetic polarities appear throughout the atmospheric columns.

In summary, similarities across these clusters of wide profiles are seen in the temperature structures, and in part in the velocities. All the clusters show a negative temperature gradient with height, with different slopes for different clusters, and at $\log\tau_{5000} \approx -1$ they have above average temperatures. The hottest atmospheres, with the weakest temperature gradients, are correlated with the strongest vertical velocities and vertical magnetic field strengths. On the other hand, the colder atmospheres tend to have weak velocities and field strengths throughout the considered regions. A significant finding is that some of the widest synthetic profiles occur in the absence of significant vertical velocities.

\subsubsection{Narrow profiles}
\label{sec:narrow_profiles}

In Fig. \ref{fig:narrow_atmos_means}, we treat four clusters with some of the narrowest profiles from each simulation. To wit, we show clusters \#32, \#57, \#69, \#75 from \textit{ch012023} (Fig. \ref{fig:ch012_unscaled_undeg_clusters_km}); clusters \#42, \#43, \#90, \#99 from \textit{nw012023} (Fig. \ref{fig:nw012_unscaled_undeg_clusters_km}); clusters \#8, \#29, \#55, \#88 from \textit{nw072100} (Fig. \ref{fig:nw072_unscaled_undeg_clusters_km}).

The \textit{ch012023} case shows similarities across all our four selected clusters. Most intriguing is the temperature, which increases $\log\tau_{5000} \approx -1$ to a local maximum around $\log\tau_{5000} \approx -3$ before sinking to a minimum around $\log\tau_{5000} \approx -5$. There is some variation in the exact height of the maximum, both between and within the clusters, but the general tendency is shared among the vast majority of the individual profiles, and stands in contrast to what we see in the other cluster `families'. Though it is not seen in the average quantities, a fraction of the profiles in these clusters do not follow the average decrease after reaching the maximum, but continue as hot streaks of fairly constant temperature all the way up to $\log\tau_{5000} \approx -5$ as seen for cluster \#75 in Fig. \ref{fig:ch012_example_31_41_70_56}.  The vertical velocities are generally low (absolute values of $< 2.5$ km s$^{-1}$), with no obvious structure. Likewise, the vertical magnetic field is very weak (absolute values of $< 2.5$ mT).

The \textit{nw012023} case reveals two distinct behaviors. Clusters \#42 and \#43 show the same sort of structure in temperature as the narrow clusters we considered for the \textit{ch012023} simulation, namely a colder layer at the bottom going to a hotter layer above before decreasing again towards the core forming heights. However, the vertical velocities tend to be slightly stronger (absolute values of $\sim 2.5$ km s$^{-1}$) and are predominantly downflowing, with occasional upflows; furthermore there are more instances of the hot streaks which do not significantly decrease in temperature from the maximum in these clusters compared to those in the previous simulation. The vertical magnetic field components for these clusters do not seem to be particularly coherent, but they are moderately strong (absolute values of $\sim 5$ mT) in the heights below. 

Somewhat different is the structure for clusters \#90 and \#99, which show consistently high temperatures throughout the line forming region, with only minor changes as a function of height. These are clusters are correlated with strong downflows ($v_z < -5$ km s$^{-1}$), and quite strong vertical magnetic field strengths (absolute values of $> 10$ mT).

A similar story is repeated for the \textit{nw072100} simulation. These are the narrowest profiles we have looked at, and the temperature structure is quite similar to the two distinct types seen for the \textit{nw012023} simulation. For the flatter high temperature cases here (clusters \#55 and \#88 from Fig. \ref{fig:nw072_unscaled_undeg_clusters_km}), the vertical velocities do not get very large (absolute values of $< 2.5$ km s$^{-1}$) before reaching a height in excess of $\log\tau_{5000} \approx -5$. However, they do contrast with the nearly zero vertical velocities for the two clusters (\#8 and \#29) with the lower starting temperatures. As in the \textit{nw012023} case there is a rise to a maximum in the temperature before it falls off again. This rise to a localized temperature maximum is not as strong in cluster \#8 as in \#29; however, it is more clearly seen when looking at the individual profiles in a manner similar to Fig. \ref{fig:ch012_example_31_41_70_56} than what is apparent from the averages shown in Fig. \ref{fig:narrow_atmos_means}. Also in this case do we find that the narrow profiles with high and consistent temperatures correlate with stronger vertical magnetic field strengths (absolute values $> 10$ mT), while the cooler atmospheres are associated with weaker vertical field strengths (absolute values of $< 2.5$ mT).

In summary, it appears that the key difference between the narrow and wide profile clusters we have examined lies in the temperature structures. The wide profiles have clear negative temperature gradients with increasing height, while the narrow profiles actually tend to have either quite flat temperatures, or an increase to a local maximum followed by a decrease.

\subsubsection{Shallow profiles}
\label{sec:shallow_profiles}

In Fig. \ref{fig:shallow_atmos_means}, we look at four of the shallower clusters from each simulation. These are clusters \#4, \#8, \#11, \#30 from \textit{ch012023} (Fig. \ref{fig:ch012_unscaled_undeg_clusters_km}); clusters \#4, \#8, \#22, \#24 from \textit{nw012023} (Fig. \ref{fig:nw012_unscaled_undeg_clusters_km}); clusters \#16, \#34, \#36, \#47 from \textit{nw072100} (Fig. \ref{fig:nw072_unscaled_undeg_clusters_km}).

In terms of the mean profiles, these clusters are some of the ones most similar to the observed clusters such as \#63 and \#81 in Fig. \ref{fig:obs250_unscaled_clusters_km}. However, it should be noted that there is quite a lot of variance within these clusters, which becomes evident when looking at the Stokes $I$ profiles for all the cluster members simultaneously. Even so, the amplitudes of the variations are not very large, and they retain the defining characteristic of being shallow, with small differences in intensity from the line wings to the line core. In all three simulations the temperatures tend to be lower than average across most of the formation region. On average the temperatures tend to decrease with increasing height, but there are a number of profiles that correspond to both extended and localized temperature enhancements in the range $-5 < \log\tau_{5000} < -3$. These temperature enhancements tend to correlate with some weak intensity enhancements around the transition from line wings to line core. In all three simulations the vertical velocities tend to be weak or moderate throughout the line forming region (absolute values of $< 2.5$ km s$^{-1}$), with the notable exception that cluster \#36 from the \textit{nw072100} simulation has strong downflows ($v_z < -5$ km s$^{-1}$)in the range $-5.5 < \log\tau_{5000} < -4.5$ corresponding to the evident redshift of the core. Similarly, the vertical magnetic field components tend to be rather weak in all cases (absolute values of $< 5$ mT), though there are some stronger fields (absolute values of $\sim 5$ mT) of both polarities present in the two colder clusters (\#34 and \#47 from \textit{nw072100}).

\subsubsection{Profiles with smooth transition to line core}
\label{sec:least_curved_profiles}

Finally, in Fig. \ref{fig:leastcurved_atmos_means} we investigate four of the clusters showing the gentlest (most gradual) transition from line wing to line core from each simulation. These are clusters \#2, \#7, \#9, \#23 from \textit{ch012023} (Fig. \ref{fig:ch012_unscaled_undeg_clusters_km}); clusters \#3, \#5, \#7, \#22 from \textit{nw012023} (Fig. \ref{fig:nw012_unscaled_undeg_clusters_km}); clusters \#13, \#14, \#20, \#45 from \textit{nw072100} (Fig. \ref{fig:nw072_unscaled_undeg_clusters_km}). These are a fairly common type of profile, comprising about 9\% of all profiles in \textit{ch012023} and \textit{nw012023}, and about 7\% in \textit{nw072100}. 


 In all cases, the mean temperature gradient of the cluster atmospheres is steeper than that of the full simulation box. As before, there are occasional instances of localized temperature enhancements (particularly for the \textit{nw012023} clusters), but they are infrequent. The clusters from the \textit{ch012023} and \textit{nw072100} simulations all have very weak vertical velocities (absolute values of $\ll 2.5$ km s$^{-1}$) all the way up to $\log\tau_{5000} \approx -5$, with the exception of \#45 from \textit{nw072100} where stronger downflows ($v_z < -2.5$ km s$^{-1}$) appear coincidentally with the temperature enhancements around $-5 < \log\tau_{5000} < -4.5$. The vertical velocities in the \textit{nw012023} clusters are not very strong, but they are consistently downflowing and somewhat larger ($v_z > -5$ km s$^{-1}$) than for the other two simulations in the heights below $-5 < \log\tau_{5000}$. For all three simulations the vertical magnetic field components appear generally quite weak (absolute values of $< 5$ mT), though the \textit{nw012023} clusters have some moderately strong fields (absolute values of $\geq 5$ mT) of both polarities interspersed among the members of the considered clusters.

\section{Discussion}
\label{sec:discussion}

We find that, while the mean profiles on the whole correspond decently well between observations and simulations, there are important differences between the observed and synthetic line profiles. Chief among them is the tendency for the synthetic profiles to be narrower than the observed profiles, both individually and in mean. This finding echoes several previous studies \citep[e.g.][]{2009ApJ...694L.128L, 2012A&A...543A..34D, 2013ApJ...764L..11D,2016ApJ...830L..30D,2016ApJ...826L..10S, 2018A&A...619A..60J} that have investigated the correspondence between synthetic and observed \ion{Ca}{ii} 854.2 nm spectra. Another key difference is the tendency for synthetic spectra to display a sharper transition from the wing to the core (a sharp `knee'); this is also seen in the results of those previous studies, but not much commented on.

In those previous studies, the discrepancy between observed and synthetic line widths is often ascribed to the effects of numerical resolution in the simulation causing less small-scale dynamics and heating. Other possible contributions have been suggested, for example  \citet{2015ApJ...809L..30C} demonstrate how the temperature profile of the atmosphere can affect \ion{Mg}{ii} profile shapes, and \citet{2013ApJ...764...40C} show how lines can be broadened from temporal averaging. The effects of resolution are certainly an important element of the explanation, as we found that our 100~km resolution simulation has both a narrower mean profile and several clusters of far narrower profiles than the two 23~km resolution simulations, even though the 100~km resolution was the most vigorous and dynamic of our simulations. However, we still see the same difference in the shape of the `knee' in both mean spectra of our 23~km resolution simulations as well as in several of their clusters, which indicates that resolution is not the only issue.

In many cases, an ad-hoc microturbulence is added in the spectral synthesis to account for such missing small-scale dynamics and improve the fit between observed and synthetic profiles. For instance, \citet{2012A&A...543A..34D} found that they needed a microturbulence of $3\; \mathrm{km\, s}^{-1}$ in order to broaden their synthetic profiles to be comparable in width to the observations of \citet{2009A&A...503..577C}. Yet, as can be seen in Fig. 2 of \citet{2012A&A...543A..34D}, the broadening due to microturbulence does not fix the sharp `knee'. Although not shown here, we repeated some of our clustering experiments after adding microturbulence in different amounts. We find the same behaviour persisting throughout the clusters, namely that the `knee' remains sharper for the synthetic profiles than the observed profiles, and adding microturbulence makes the profile wider only near the line core, resulting in profile shapes that still have a `knee' that is sharper than observed. 

Our investigation into the atmospheric structure for the different clusters revealed that the key parameter in setting spectral shapes that resemble the observations is the temperature stratification, in particular the temperature gradient in the region $-3 \lesssim \log\tau_{5000} \lesssim -1$. A stronger temperature gradient in this region typically leads to broader profiles and a gentle wing-to-core transition. This can be seen in our groups of clusters showing the ``shallow'' profiles (Fig.~\ref{fig:shallow_atmos_means}), or smoothest transition from wing to core (Fig.~\ref{fig:leastcurved_atmos_means}), all of which have a stronger gradient than the average of each simulation. Interestingly, in their Fig.~3, \citet{2010ApJ...722.1416M} show a comparison of the \ion{Ca}{ii} 854.2 nm intensity profiles from the semi-empirical FALC and M-CO (also known as FALX) atmospheres where the M-CO result displays an appreciably smoother `knee'. Compared to FALC, the M-CO atmosphere has a temperature gradient that persists into a cooler temperature minimum at 1~Mm (corresponding to $ \log\tau_{5000} \approx -5.2$, compared to the 500~km (corresponding to $\log\tau_{5000} \approx -3.5$ of FALC (the M-CO synthetic line core has other shortcomings, such as a saturated core, that do not resemble observations). Another revealing aspect from our present work is that the broadest synthetic profiles, associated with clusters that have strong velocities (Fig.~\ref{fig:wide_atmos_means}), are broad but still have a sharp `knee'. This strongly suggests that turbulent broadening, whether by real atmospheric motions or by adding microturbulence, does not lead to the spectral shapes of observed quiet Sun profiles. 

Our analysis presents a new way to quantify differences in spatially-resolved spectra that go beyond averages and use more information than simpler line properties such as shifts or widths. This can be used as a stringent test when comparing observations and simulations, but also to learn about the formation of typical spectral shapes. A key result is that the simulations we tested show much larger variations that the observations. The shapes of the most common spectral clusters are richer for the synthetic profiles, where strongly asymmetric and emission profiles are much more common than in observations. The reason for this additional variation is not yet understood. It is also puzzling that a lower activity level in the simulations does not seem to lead to less variation in spectral shapes. But our comparison with the most active simulation we had available  (\textit{nw072100}) is not completely fair because it has a coarser spatial resolution and an area 36 times larger than the less active simulations. Another aspect that could affect the comparison and appearance of synthetic profiles is that our synthetic spectra are an instant snapshot, while the observations were a scan through wavelength that took a few seconds to complete. We were unable to test this scanning effect with synthetic spectra at this point, but \citet{2023A&A...669A..78S} find that this can affect the profiles, in particular in areas with magnetic features.

\section{Conclusions}
\label{sec:conclusion}

We performed a comparative clustering analysis of the synthetic \ion{Ca}{ii} 854.2 nm profiles from three simulations of varying magnetic activity alongside quiet Sun observations of the same spectral line. We found that the clusters retrieved from the observations and the simulations show similarities, but also significant differences that persist even after the synthetic profiles have been spatially and spectrally degraded to match the observational conditions. 

The most obvious difference was the that the observed profiles were generally wider than the synthetic. However, we also see a tendency for the synthetic absorption profiles to display steeper transitions from the line wings to the core, while the observed absorption profiles in general had gentler transitions. Another key difference was that the observations contain far fewer profiles with emission peaks, strongly asymmetric profiles, or complex line profiles than we see in the synthetic spectra. This is a possible indication that even our most quiet simulation is more dynamic than the observed region.

When we compared the synthetic profiles from the simulations with each other, we find that the largest difference between the retrieved clusters was that the more active simulations produced larger intensity differences and more variance within the clusters, but mostly the same profile shapes appeared in all of the simulations. A specific type of profiles with broad emission peaks is found only in the most vigorous simulation. Additionally, that simulation also had a significantly lower horizontal resolution than the other two, and produced the narrowest absorption profiles.

Furthermore, we investigated the atmospheric structure for a few selected clusters of absorption profiles: the widest, the narrowest, the shallowest, and profiles with the least steep transitions from line wing to core. We find that the strongest correlations and differences between these types of clusters appear in the temperature stratifications. Both the shallow and least steep line to wing clusters came from atmospheres with lower than average temperatures and their temperatures typically followed a monotonous negative gradient with increasing heights, with occasional occurrences of localized heating. For these clusters, the vertical velocities and vertical magnetic field components were generally quite weak throughout the line forming region. The profile shapes in these clusters are the closest to observations, which suggests that the quiet Sun has steeper temperature gradients in the range $ -3 \lesssim \log\tau_{5000} \lesssim -1$, than the simulation averages.

Concerning the clusters of wider profiles profiles, we find some clusters show weak velocities and magnetic field strengths, and other clusters show stronger flows and field strengths. The profiles with the weaker velocities tend to have more prominent temperature minima, with the temperature decreasing steadily throughout the range $ -5 \lesssim \log\tau_{5000} \lesssim -2$. The profiles associated with stronger velocities and field strengths coincide with higher and flatter temperatures, often reaching all the way up to around $ \log\tau_{5000} \approx -5$. This demonstrates that large velocities are not necessary ingredients for producing wide profiles. 

For the clusters with narrowest profiles we find that the temperature stratification took on a markedly different character compared to the other clusters we have considered. In the cases where the velocities and magnetic field strengths were weak, the temperature tends to rise to a significantly higher than average local maximum around $ \log\tau_{5000} \approx -3$ before sinking again. In the cases where the vertical velocities and magnetic field strengths took on higher values, the temperatures took on and maintain quite constant and high values throughout the whole line forming region. This shows that strong velocities alone are not sufficient to produce wide profiles.

In sum, we find indications that the temperature stratification, more so than the vertical velocities, holds the key for getting the simulations to produce synthetic \ion{Ca}{ii} 854.2 nm profiles more similar to quiet Sun observations.

\begin{acknowledgements}
We wish to thank the anonymous referee for constructive and thorough feedback, which lead to several improvements of this manuscript.
This work has been supported by the Research Council of Norway through its Centers of Excellence scheme, project number 262622. Computational resources have been provided by UNINETT Sigma2 - the National Infrastructure for High Performance Computing and Data Storage in Norway. The computations were enabled by resources provided by the Swedish National Infrastructure for Computing (SNIC) at the PDC Centre for High Performance Computing (PDC-HPC) at the Royal Institute of Technology partially funded by the Swedish Research Council through grant agreement no. 2018-05973.
The Swedish 1-m Solar Telescope is operated on the island of La Palma by the Institute for Solar Physics of Stockholm University in the Spanish Observatorio del Roque de los Muchachos of the Instituto de Astrof{\'\i}sica de Canarias.
The Institute for Solar Physics is supported by a grant for research infrastructures of national importance from the Swedish Research Council (registration number 2021-00169).
\end{acknowledgements}

\bibliography{Refs.bib}

\begin{appendix}
\section{The clusters for the degraded synthetic profiles} 
\label{sec:appendix_degraded_clusters}

\begin{figure*}
        \centering \hspace*{-5mm}
                \includegraphics[width=19.5cm]{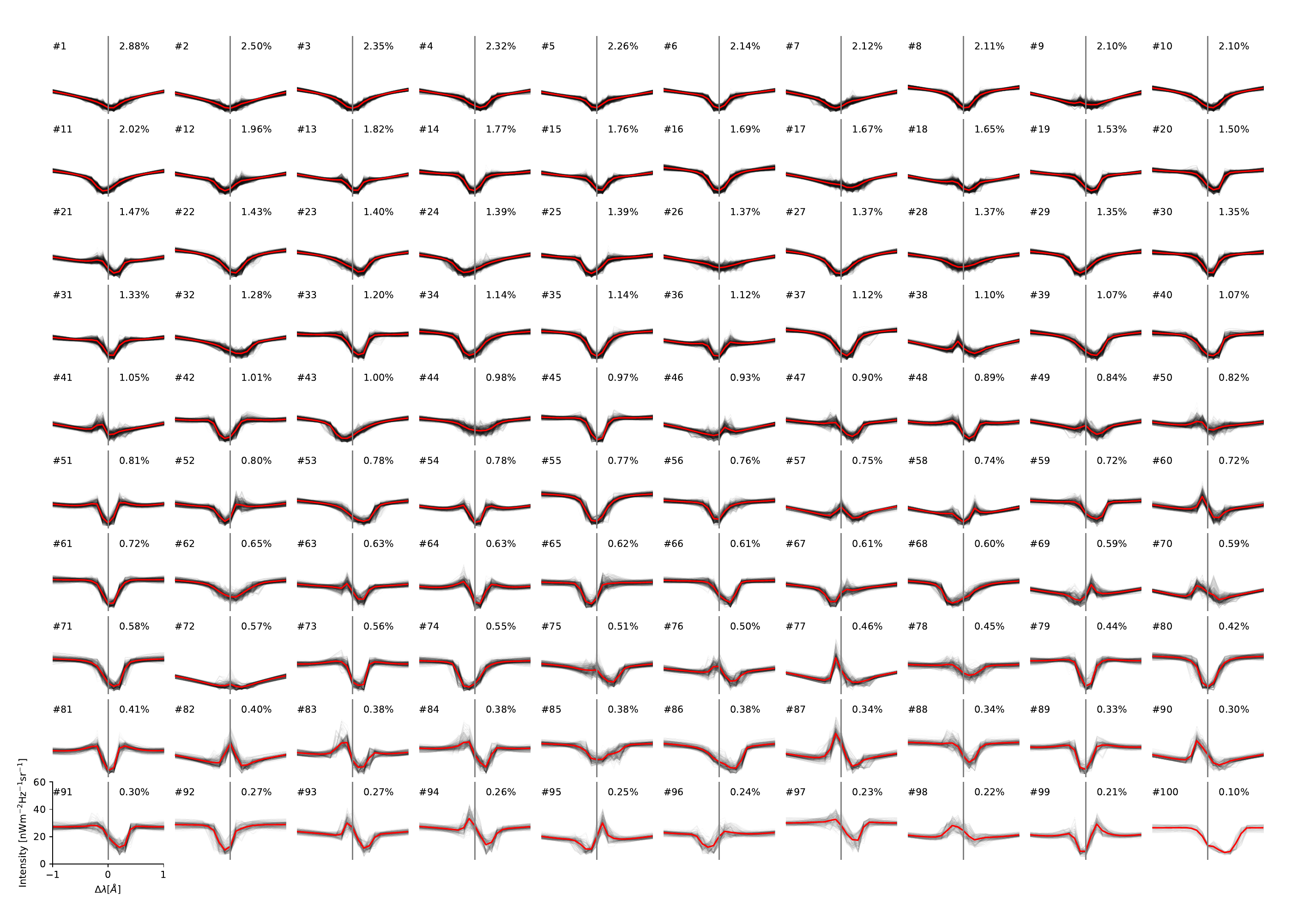}

                        \caption{$k$-means clusters of synthetic \ion{Ca}{II} 854.2~nm intensity profiles, using 100 clusters for the unnormalized and degraded profiles from \textit{ch012023}. The red line is the cluster mean, and the black lines are all the individual profiles belonging to each cluster. The grey line indicates the position of $\lambda_0$.}
                        \label{fig:ch012_unscaled_deg_clusters_km}
\end{figure*}

\begin{figure*}
        \centering \hspace*{-5mm}
                \includegraphics[width=19.5cm]{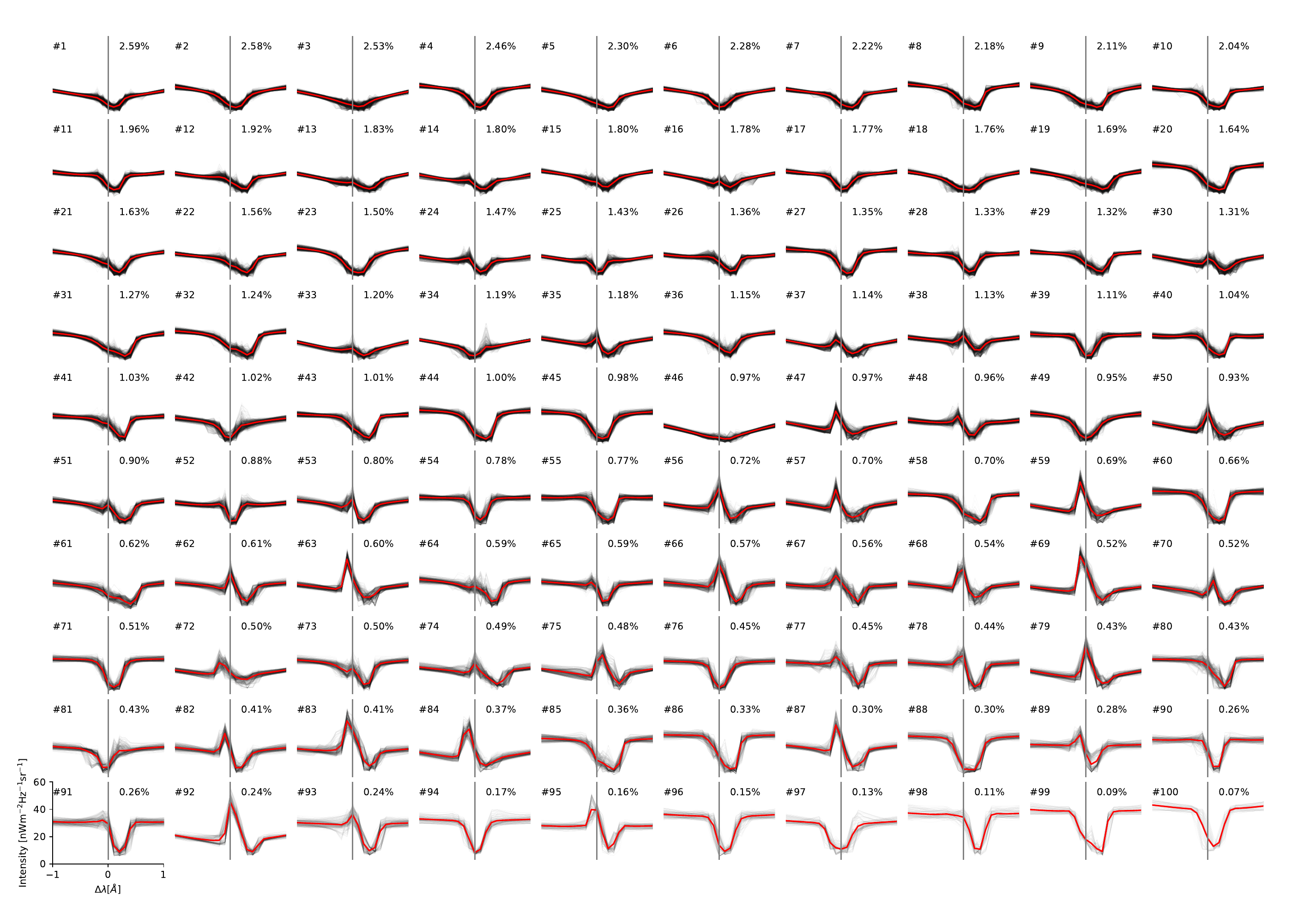}

                        \caption{$k$-means clusters of synthetic \ion{Ca}{II} 854.2~nm intensity profiles, using 100 clusters for the unnormalized and degraded profiles from \textit{nw012023}. The red line is the cluster mean, and the black lines are all the individual profiles belonging to each cluster. The grey line indicates the position of $\lambda_0$.}
                        \label{fig:nw012_unscaled_deg_clusters_km}
\end{figure*}

\begin{figure*}
        \centering \hspace*{-5mm}
                \includegraphics[width=19.5cm]{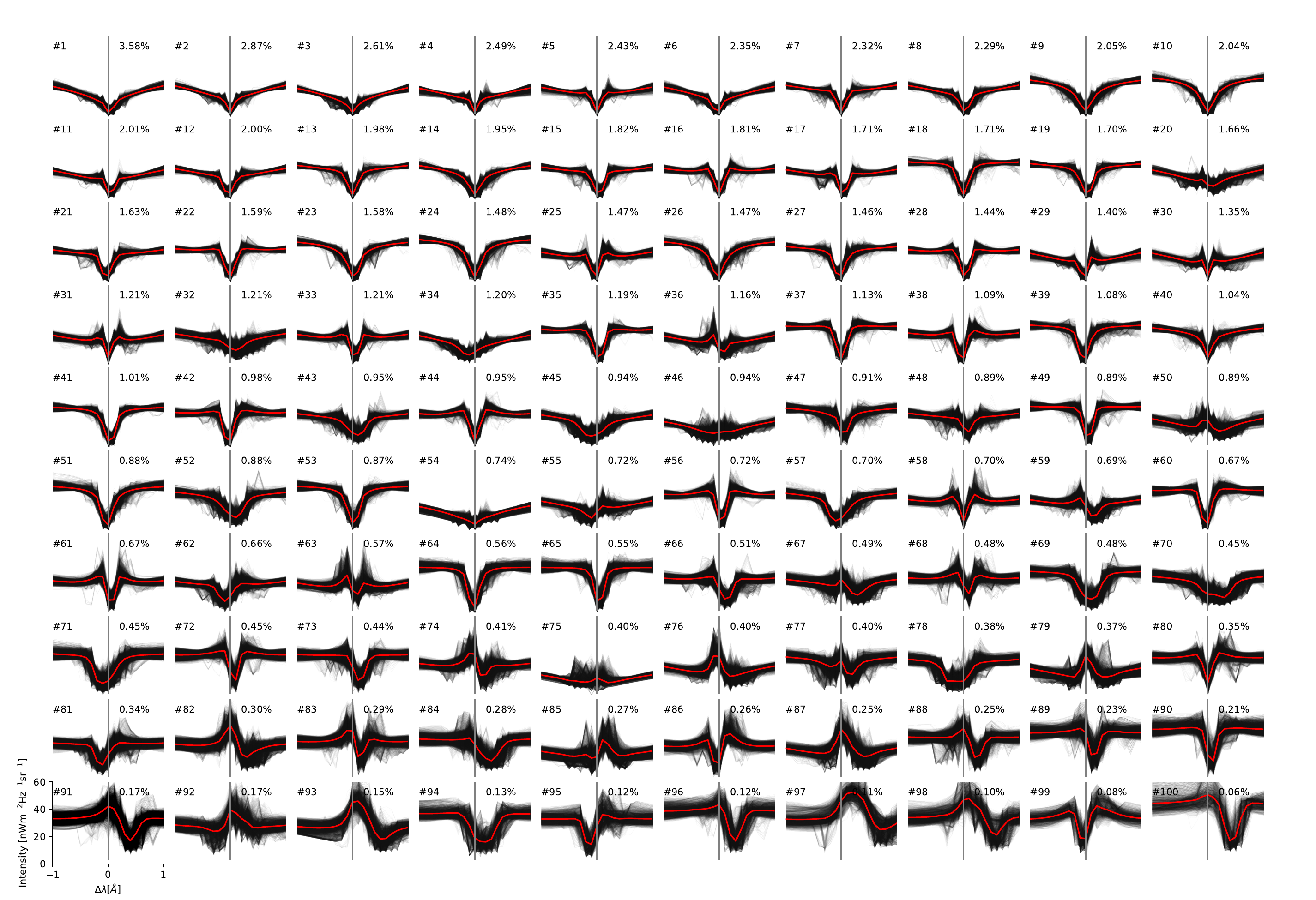}

                        \caption{$k$-means clusters of synthetic \ion{Ca}{II} 854.2~nm intensity profiles, using 100 clusters for the unnormalized and degraded profiles from \textit{nw072100}. The red line is the cluster mean, and the black lines are all the individual profiles belonging to each cluster. The grey line indicates the position of $\lambda_0$.}
                        \label{fig:nw072_unscaled_deg_clusters_km}
\end{figure*}

In Sect. \ref{sec:clusters}, we present the resulting clusters after applying $k$-means to our observed profiles and to our undegraded synthetic profiles. We performed the same type of clustering also for synthetic profiles that have been spectrally and spatially degraded and downsampled to match the observational conditions, and the results are shown in Figs. \ref{fig:ch012_unscaled_deg_clusters_km}, \ref{fig:nw012_unscaled_deg_clusters_km}, \ref{fig:nw072_unscaled_deg_clusters_km}. The net effect of the degradation is primarily manifested as a reduction in the variance of the profiles with regards to absolute intensity. However, the general shapes persist from the undegraded clusters, and the qualitative differences between the observed and synthetic line profiles remain similar. Notably, there are still far more exotic profiles present in the degraded clusters compared to the distribution that is seen in the observations.

\section{Additional figures of atmospheric structure}
\label{appendix:extra_atmos_figures}

\begin{figure*}
        \centering \hspace*{-5mm}
                \includegraphics[width=19.5cm]{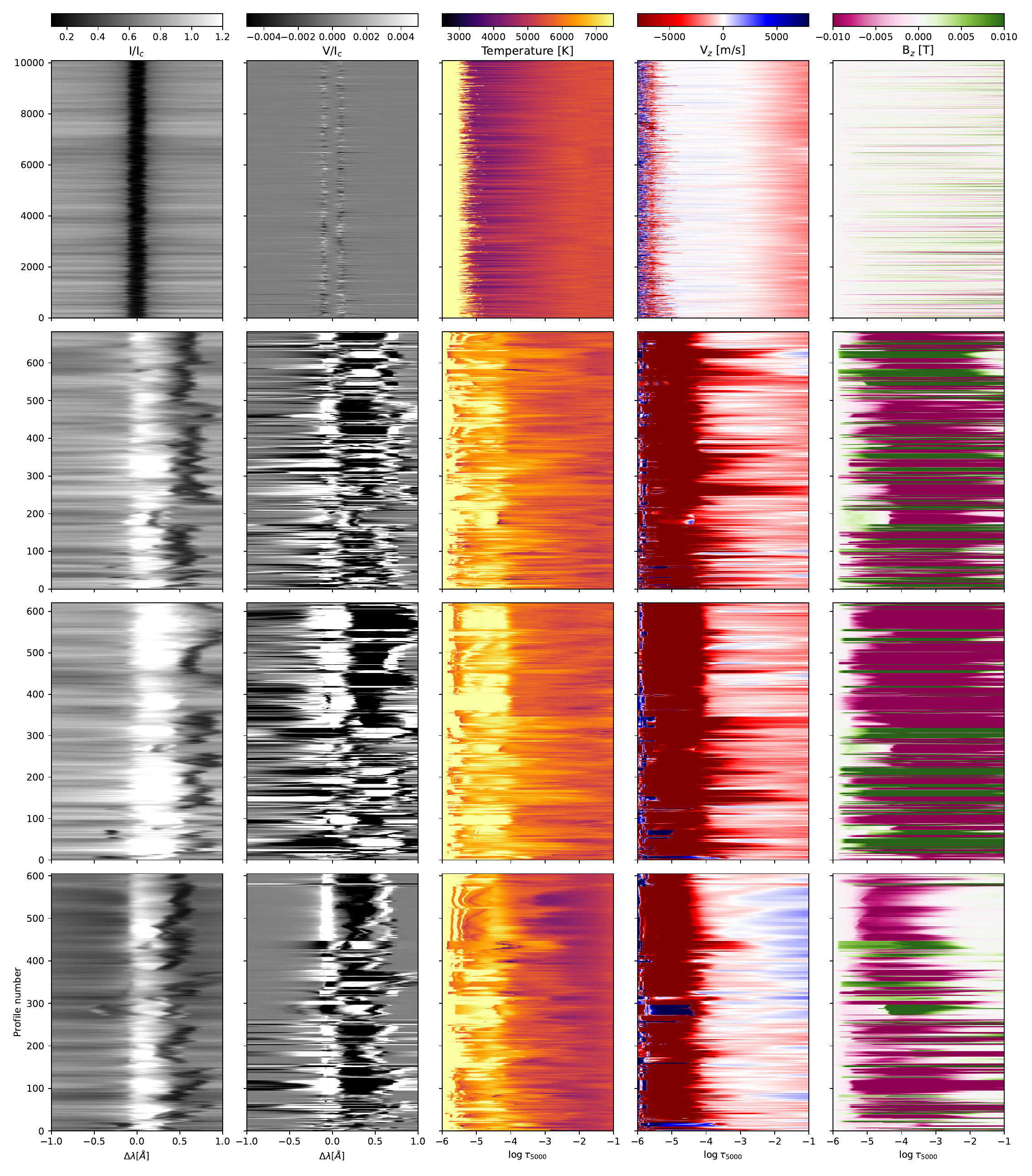}

                        \caption{Shows the atmospheric structure of clusters \#15, \#93, \#94, \#96 in Fig. \ref{fig:nw072_unscaled_undeg_clusters_km}. The top panel contains a cluster of the typical absorption profiles, while the three bottom panels are the clusters showing broad emission in \textit{nw072100}, following the same presentation as Fig. \ref{fig:ch012_example_31_41_70_56}.}
                        \label{fig:nw072_broad_emission_41_79_48_96}
\end{figure*}

In Fig.~\ref{fig:nw072_broad_emission_41_79_48_96} we show the spectra and atmospheric profiles from four clusters from the \textit{nw072100} simulation, which include shallow absorption profile and more dynamic clusters with broad emission.

\section{A few results from using $k$-Shape}
\label{appendix:some_kshapes_results}
This section shows a few examples of the clusters retrieved by using $k$-Shape on $z$-normalized spectra. Compared to the $k$-means results (where spectra were not normalized), there are more complicated shapes present in these clusters, which for the most part are `hidden' in the least constrained and most shallow clusters in the original cases shown in the main text. 

\begin{figure*}
        \centering \hspace*{-5mm}
                \includegraphics[width=19.5cm]{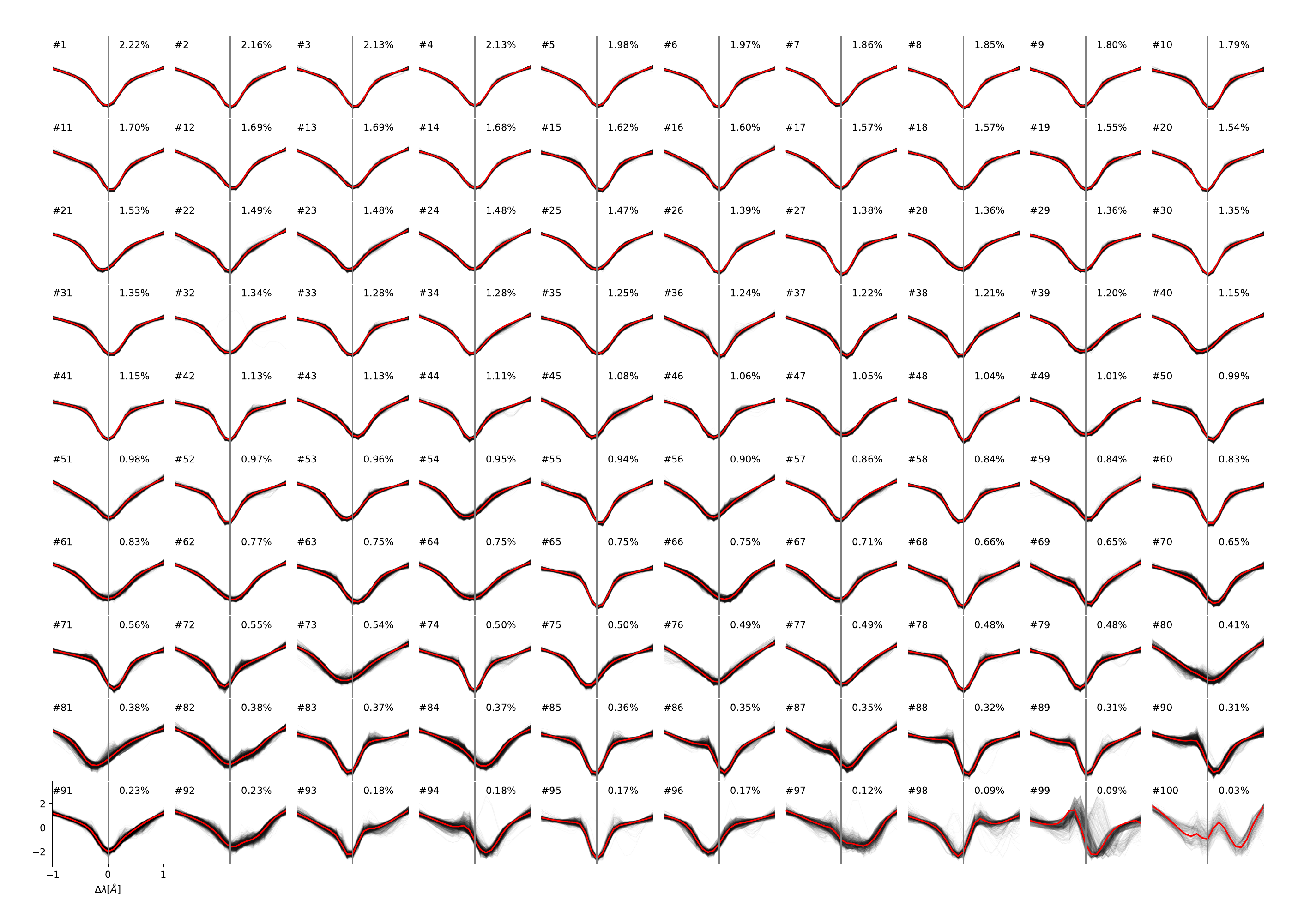}

                        \caption{Showing the clusters found with $k=100$ clusters using $k$-Shape with 10 re-initialization on the $z$-normalized observed spectra.}
                        \label{fig:obs250_clusters_ks_init10}
\end{figure*}

As examples, Fig. \ref{fig:obs250_clusters_ks_init10} has cluster \#28, which primarily corresponds to \#81 and \#97 in Fig. \ref{fig:obs250_unscaled_clusters_km}; \#25 mostly corresponds to \#89, \#97, \#99; \#94 mostly goes into \#90, \#97, \#100; \#16 for the most part splits into \#17, \#45, \#60, \#63, \#81.

\begin{figure*}
        \centering \hspace*{-5mm}
                \includegraphics[width=19.5cm]{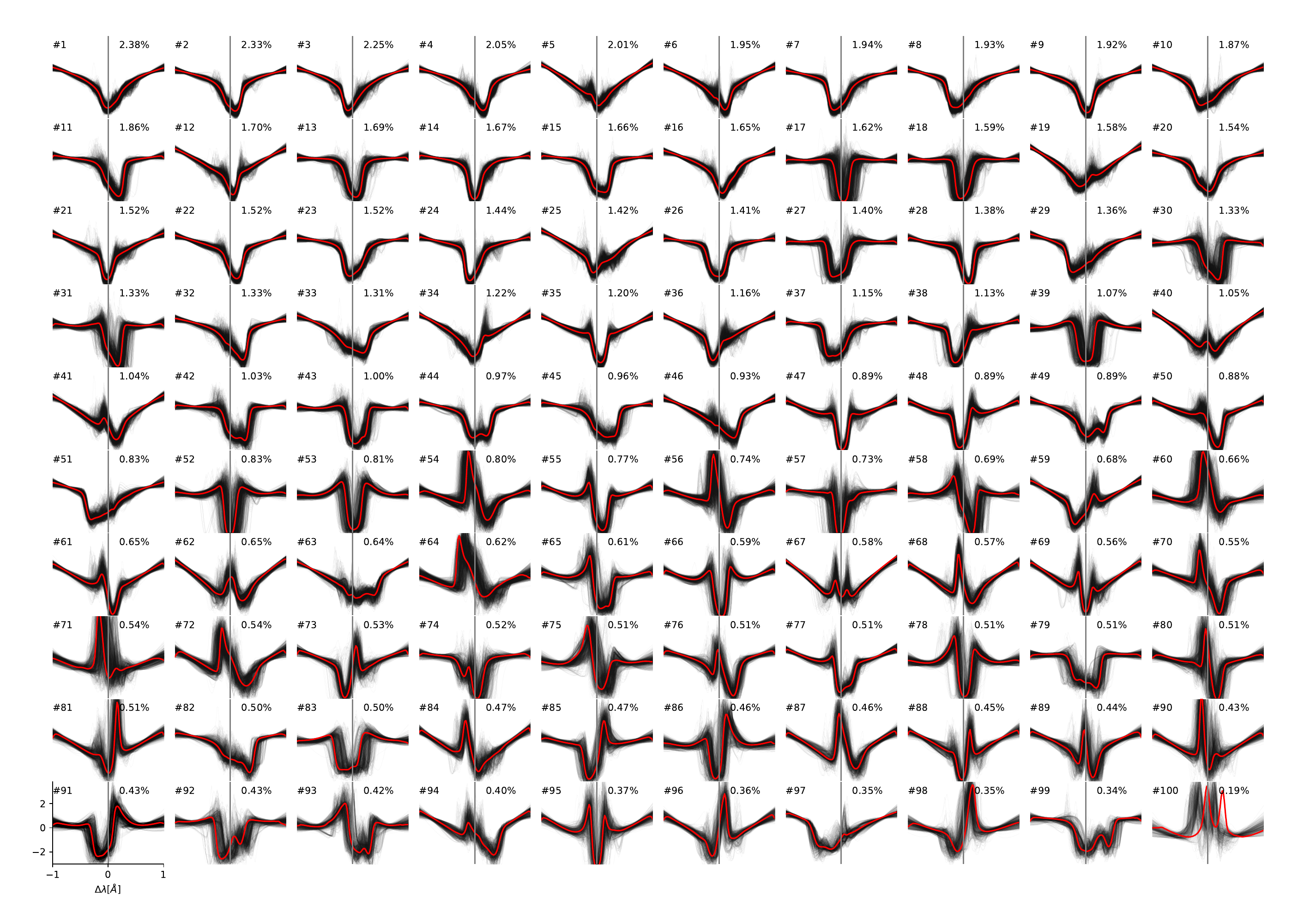}

                        \caption{Showing the clusters found with $k=100$ clusters using $k$-Shape with a single initialization on the $z$-normalized undegraded spectra for the \textit{ch012023} simulation.}
                        \label{fig:ch012_undeg_clusters_ks}
\end{figure*}

\begin{figure*}
        \centering \hspace*{-5mm}
                \includegraphics[width=19.5cm]{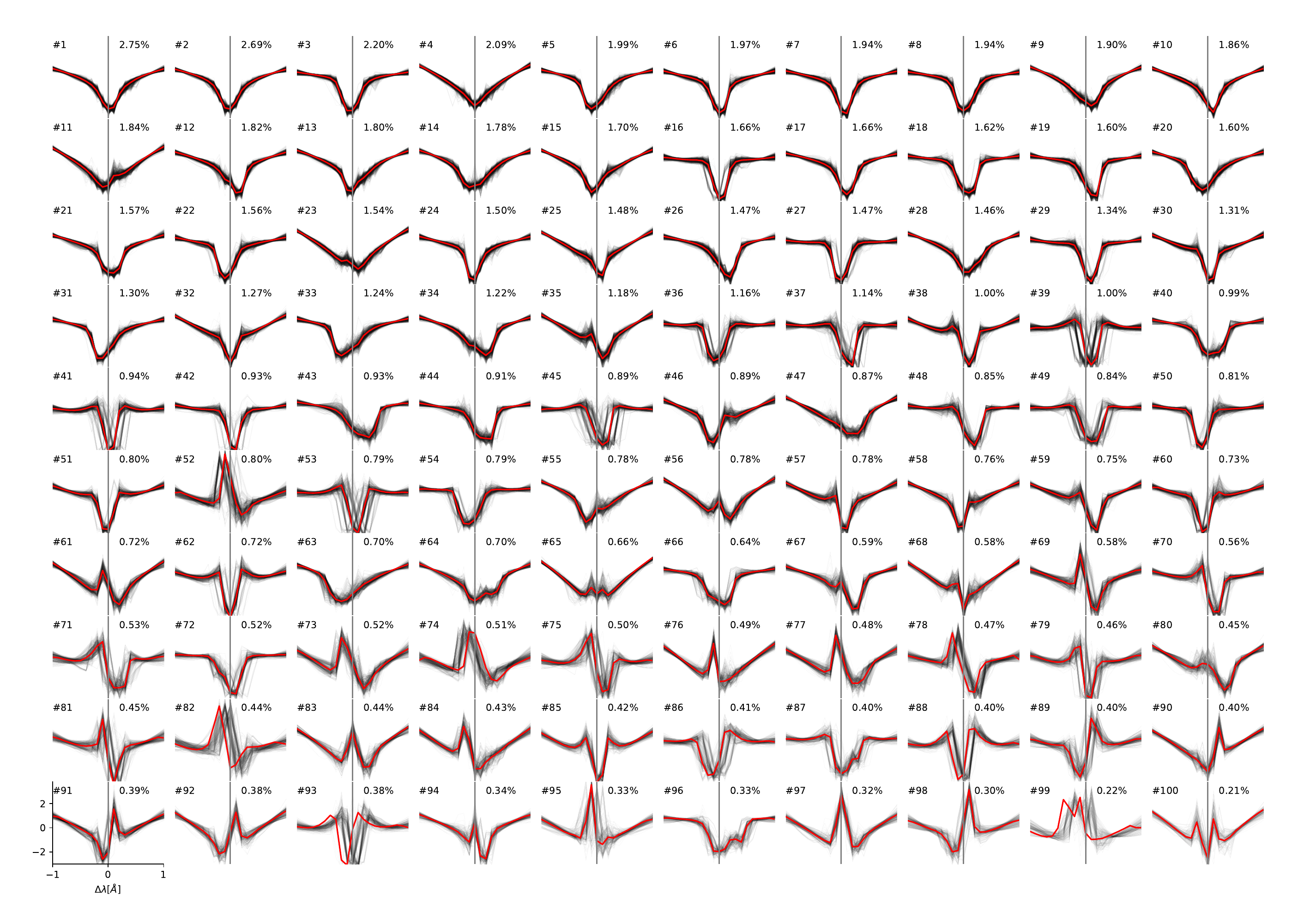}

                        \caption{Showing the clusters found with $k=100$ clusters using $k$-Shape with a single initialization on the $z$-normalized degraded spectra for the \textit{ch012023} simulation.}
                        \label{fig:ch012_deg_clusters_ks}
\end{figure*}

Fig. \ref{fig:ch012_undeg_clusters_ks} has unique clusters such as \#19, \#46, \#64, \#75, \#77, \#85, \#100  which do not appear with corresponding centroids in Fig. \ref{fig:ch012_unscaled_undeg_clusters_km}. \#19 is pretty well spread but the larger receivers are \#21, \#44, \#50, \#55, \#73, \#92; \#46 goes mostly into \#13; \#64 goes into many different clusters, but the larger receivers are \#3, \#7, \#28, \#30, \#33, \#34; \#75 goes primarily into \#81, \#99; \#77 goes mostly into \#30, \#35, \#55, \#73;  \#85 is also quite spread with the most frequent destinations being \#14, \#25, \#27, \#55; \#100 goes mostly into \#13, \#29, \#62. Even after degradation, the same complex shapes still appear as seen in Fig. \ref{fig:ch012_deg_clusters_ks}. We have not performed more than a cursory inspection for the other synthesized spectra, but at a preliminary glance similar trends seem to hold for these as well.

\end{appendix}

\end{document}